\documentclass[structabstract]{aa}  
\usepackage{graphicx}
\usepackage{txfonts}
\usepackage{graphicx}
\usepackage{natbib}
\usepackage{pifont}
\newcommand\bk{\bar\kappa}
\newcommand\gc{\gamma_{\rm c}}
\newcommand\gw{\gamma_{\rm w}}
\newcommand\gt{\gamma_{\rm g}}
\newcommand\Pc{P_{\rm c}}
\newcommand\Pg{P_{\rm g}}
\newcommand\Pw{P_{\rm w}}
\newcommand\Ec{E_{\rm c}}
\newcommand\Eg{E_{\rm g}}
\newcommand\Ew{E_{\rm w}}
\newcommand\vA{v_{\rm A}}
\newcommand\us{u_{\rm s}}
\newcommand\uw{u_{\rm w}}
\newcommand\cs{c_{\rm s}}
\newcommand\bvA{{\bf v}_{\rm A}}
\newcommand\dBs{\langle(\delta{\bf B})^2\rangle}
\newcommand\tacc{\tau_{\rm acc}}
\newcommand\pmax{p_{\rm max}}
\newcommand\urel{u^{\rm rel}}
\def\pop#1#2{\frac{\partial#1}{\partial#2}}
\def\diver#1{\nabla\!\cdot\!#1}
\newcommand\lenq{l_{\rm q}}
\newcommand\sgu{(\nabla{\bf u})}
\newcommand\DV{\Delta V}
\newcommand\Zdri{\frac{1}{3}}

\newcommand\Msol{M_\odot}
\begin{document}
\title{Time-dependent galactic winds}
\subtitle{I. Structure and evolution of galactic outflows accompanied by cosmic ray 
          acceleration}

\author{E. A. Dorfi      \inst{1}
   \and D. Breitschwerdt \inst{2} }
   
\offprints{E.A. Dorfi}

\institute{Universit\"at Wien, Institut f\"ur Astronomie,
           T\"urkenschanzstr. 17, A-1180 Wien, Austria \\
           \email{ernst.dorfi@univie.ac.at}
           \and
           Zentrum f\"ur Astronomie und Astrophysik, Technische Universit\"at Berlin,
           Hardenbergstra{\ss}e 36, D-10623 Berlin, Germany \\
           \email{breitschwerdt@astro.physik.tu-berlin.de}  }

\date{Received date .......; accepted date .......}

\abstract
{Cosmic rays (CRs) are transported out of the galaxy by diffusion and advection due to streaming 
along magnetic field lines and resonant scattering off self-excited MHD waves. Thus momentum is 
transferred to the plasma via the frozen-in waves as a mediator assisting the thermal pressure in 
driving a galactic wind.}
  {The bulk of the Galactic CRs (GCRs) are accelerated by shock waves generated in supernova remnants (SNRs), a significant fraction of which occur in OB associations on a timescale of several $10^7$ years. We examine the effect of changing boundary conditions at the base of the galactic wind due to sequential SN explosions on the outflow. Thus pressure waves will steepen into shock waves leading to \emph{in situ post-acceleration of GCRs}.}
  {We performed hydrodynamical simulations of galactic winds in flux tube geometry appropriate for disk galaxies, describing the CR diffusive-advective transport in a hydrodynamical fashion (by taking appropriate moments of the Fokker-Planck equation) along with the energy exchange with self-generated MHD waves.}
  {Our time-dependent CR hydrodynamic simulations confirm the existence of time 
asymptotic outflow solutions (for constant boundary conditions), which are in excellent 
the agreement with the steady state galactic wind solutions  
described by Breitschwerdt et al. (1991, A\&A 245, 79). 
It is also found that high-energy particles escaping from the Galaxy
and having a power-law distribution in energy ($\propto E^{-2.7}$)
similar to the Milky Way with an upper energy  cut-off at $\sim
10^{15}$ eV are subjected to efficient and rapid post-SNR
acceleration in the lower galactic halo up to energies of $10^{17}
- 10^{18}$ eV by multiple shock waves propagating through the halo. 
The particles can gain energy within less than $3\,$kpc from the
galactic plane corresponding to flow times less than 
$5\!\cdot\! 10^6\,$years. Since particles are advected downstream of the shocks, i.e. towards the galactic disk, they should be easily observable, and their flux should be fairly isotropic.}
  {The mechanism described here offers a natural and elegant solution to
explain the power-law distribution of CRs between the ``knee'' and the ``ankle''.}


\maketitle

\section{Introduction}
It is now understood that galactic winds may form an important
evolutionary stage in galaxy evolution. In particular in the early
universe, winds may be held responsible for polluting the
intergalactic medium (IGM) with metals, possibly synthesized in
Population III stars \citep[e.g.,][]{Loew:01}. The discovery of
Lyman-Break galaxies \citep[e.g.,][]{Steid:96, AdSt:00} has given
much support to this idea. Evidence of winds at high redshift has been
reported, e.g.,\ by \citet{Daw:02}, who serendipitously discovered a
strong asymmetric Ly$_\alpha$ emission line in a faint compact
galaxy at $z=5.189$, indicating an outflow velocity in the range of
$320 -360 \, {\rm km} \, {\rm s}^{-1}$. 

In a recent study, it has been shown that
many star-forming galaxies at $2 < z < 3$ exhibit galactic outflows \citep[]{Steid:10}. 
Apart from metal enrichment
of the IGM, galactic winds are also thought to preheat the gas. The
X-ray luminosity versus temperature relationship ($L_X -T$) shows a
deviation from a power-law when going from massive to poor clusters of
galaxies and groups. This has been interpreted in terms of a
so-called ``entropy floor'' that dominates the gravitational heating
of the cluster as the potential wells become shallower
\citep{Pon:99}. In other words, the preheating of the IGM by
starburst driven galactic winds starts to dominate the release of
gravitational energy. In addition, the fairly high metallicties of
$Z \sim 0.3 \, Z_\odot$ \citep[e.g.,][]{Mol:99,Kik:99} found in the
intracluster gas may be explained by ejection of metals from
galaxies by massive winds during starburst and also more quiescent phases \citep{Kap:06}. 

In the
local universe, evidence for galactic winds mainly stems from
observations of starburst galaxies, such as \object{M82} or
\object{NGC 253}. In these more evolved objects, starbursts are
thought to be triggered by a substantial disturbance of the
gravitational potential, e.g.~by interaction with a companion
galaxy. Finally, dwarf galaxies are prone to galactic winds because
of their shallow gravitational potential well and their
correspondingly low escape speeds, easily reached by supernova
heated gas of a few $10^6$ K. According to
Bernoulli's equation, the typical outflow velocity, if all thermal
is converted into kinetic energy, is on the order of $v \simeq 250$ km/s.
In contrast, it is
still widely believed that late-type spirals like the Milky Way
should not exhibit winds, simply because the total thermal pressure
in the disk is not sufficient to enable break-out through the
extended gaseous HI layer and drive an outflow. 

The situation though
may be different in very active star forming regions, where
superbubbles {\em locally\/} inject  a substantial amount of energy
enabling hot gas to reach considerable heights \citep{AB:04, AB:05}. A
related feature was observed in H$_{\alpha}$ by
\citet{Rey:01} with WHAM. They find a giant loop extending out to 
$\sim 1300$ pc perpendicular to the galactic plane. The base of the loop
is spatially and kinematically connected to the HI chimney found by
\citet{Nor:96}, which is generated by the W4 star cluster. As more
energy is deposited into this system by stellar winds and successive
supernova explosions, we can expect the loop (the inner
ionized part of the shell) to expand further and fragment due to
Rayleigh-Taylor instabilities. The released hot plasma will then
form the base of a galactic fountain, in which the gas may travel
several kpc above the plane, but owing adiabatic and radiative
cooling, thermal instabilities will progress and condensation of
clouds (which are still gravitationally bound to the galaxy) is
supposed to occur \citep{Bre:80,Kahn,deA:00}. Only
if the combined thermal energy of breaking-out superbubbles
exceeds the gravitational potential, a thermal galactic wind
occur. In normal spirals this will only happen {\em locally}, and
more extended winds are only expected in starburst regions like in
M82 or NGC\,253.

However, as emphasized by many authors in the past 
\citep[e.g.,][]{Axf:81,Ip:75,DB91, Ev:08}, the above
argument ignores the effect of cosmic rays (CRs), which have an
energy density that is roughly in equipartition with the thermal gas and
the magnetic energy densities. The fact that CRs escape from the
Galaxy after a few tens of million years on average, setting up a
pressure gradient {\em pointing away} from the disk, gives rise to
the resonant excitation of MHD waves via the so-called CR
streaming instability \citep[e.g.,][]{Kul:Pea}. Thus CRs are
coupled to the thermal plasma by scattering off the frozen-in
waves, thereby helping to push the plasma against the
gravitational pull. As a result, CR transport, which is mainly
{\em diffusive in the disk}, will predominantly become {\em advective
in the Galactic halo}. Thus even in case of the Milky Way, a
steady state wind flow could be set up with a global mass loss
rate of $\sim 0.3 \, {\rm M}_\odot \, {\rm yr}^{-1}$
and a low outflow base velocity of $\sim 10$ km/s \citep{DB93}. In fact, it has been
shown that the diffuse broad-band soft X-ray background can be
explained all the way from $0.3\,$keV to $1.5\,$keV by a superposition of
emission from the Local Bubble and by a wind from the Galactic halo
with both plasmas being in a state of nonequilibrium ionization
\citep{BS:94,BS:99}. Recently, \citet{Ev:08} have shown that a kiloparsec 
scale galactic wind of the Milky Way gives the best explanation for the 
longitude averaged soft X-ray emission near 0.65 and 0.85 keV observed 
with ROSAT PSPC. Their model is further constrained by simultaneously fitting the observed 
radio synchrotron emission, resulting in a higher pressure and smaller disk surface 
area wind \citep{Ev:10}. 

Common to all investigations is the energy source due to more or
less frequent supernova (SN) explosions, as well as to stellar winds,
which contribute a minor fraction, and only during the initial
phases, to the momentum and energy input, because only very massive
O stars have sizable winds. For
high stellar birth rates, therefore we can parameterize the results by the star
formation rate (SFR) which relates the newly formed stars to the SN
rate. In the case of low activity, the overall SN rate (types I and
II) determines the properties of the {\em global } galactic wind.
Assuming supernovae to be the sources of cosmic rays, we indeed expect CRs to
accompany outflows from the disk into the halo, unless their
coupling is very weak, and diffusion is the only mode of transport.
However, as CR observations suggest, the diffusion halo only extends
$\sim 1 - 3$ kpc perpendicular to the disk \citep[see][]{Br:02}. We
therefore argue that CRs will always take part in the
acceleration of an outflow, and although their effect may be small
for starburst galaxies, it will dominate in the case of
normal spirals.

On the other hand, advection of CRs in the galactic wind flow may lead 
to reacceleration of escaping disk particles in the energy range 
$3 \times 10^{15} - 10^{17}$ eV (i.e.\ between the ``knee'' and the 
``ankle'')  by propagating compressional 
waves as has been suggested by \citet{VZ:04}. 
These waves come from the interaction between fast outflows, originating 
in regions in the disk that have been compressed by the spiral density 
shock wave, hence have generated a faster CR driven outflow (cf. Breitschwerdt et al. 
2002) and slower outflows from interarm regions. The steepening of the waves 
into moderate shocks of a sawtooth shape at large distances, where the flow is 
basically radial, leads to reaccelerating the particles close to the knee at a 
quasi-perpendicular shock (since the field is basically azimuthal there). 
Moreover, the supersonic galactic wind expands into the intergalactic medium with a given (but poorly 
known) pressure and must therefore be bounded by a termination shock, which itself 
can be the site of particle acceleration \citep{JM:85,JM:87}. It will, however, be  
difficult to observe these particles since they are advected away from the Galaxy with 
little chance to diffuse upstream to the disk. Since MHD turbulence downstream of the 
termination shock is presumably very strong, it acts in the model of \citet{VZ:04}
as a de facto reflecting boundary for the accelerated particles up to a certain energy. 
As the spiral density wave pattern is not in corotation with the disk and hence with the 
frozen-in halo field lines the associated forming halo shocks slip through the flux tubes. 

In this paper we present an alternative possibility of accelerating particles in a galactic 
wind flow beyond the knee. The basic idea is that individual star-forming regions, e.g. superbubbles 
or localized starbursts, have an energy output that is not constant in time. Moreover, their 
lifetime is typically less than $10^8$ years and therefore smaller than the wind flow time.  
This will lead to \textit{time-dependent} switch-off and switch-on effects of combined disk 
gas and CR outflows, which periodically push a piston into the gas. As we 
show in this paper, the galactic wind flow will be modulated by successive pairs of shocks  
(forward and reverse shock), where efficient particle acceleration can take place. Like the 
termination shock and unlike disk SNRs, these shocks will be long-lived, so the 
energy cut-off argument  does not apply in this case because of finite size and acceleration time. 
They are thus strong candidates for generating the CR spectrum between 
$10^{15}$ eV and $10^{18}$ eV by post-acceleration of GCRs, as we shall see. 

The paper is structured as follows. In Section~\ref{s.model} we
describe the physical and numerical requirements of our model, and
in Section~\ref{s.res} our general results from time-dependent
galactic wind simulations for the Milky Way are presented, in particular the
effect of particle diffusion and the propagation of
shock waves within galactic winds. In
Section~\ref{s.acc} we introduce the interesting new possibility
of accelerating cosmic rays by successive shock waves, which are
driven by supernova explosions in the underlying disk, and
propagate into the halo. A discussion and our conclusions are given in
Section~\ref{s.con}.

\section{Galactic wind model}
\label{s.model}

\subsection{Physical equations}
\label{s.eqn}

When employing the usual physical notation we use the following
time-dependent system of hydrodynamical equations supplemented 
by three energy density equations
for the thermal gas, $\Eg$,
the cosmic rays, $\Ec$, and the Alfv\'enic waves $\Ew$:
\begin{eqnarray}
\pop{\rho}{t} + \diver{(\rho{\bf u})} &=& 0
\label{mass}\\
\pop{\rho{\bf u}}{t} + \diver{(\rho{\bf u}\!:\!{\bf u})}
                  + \nabla(\Pg+\Pc+\Pw) & & \nonumber \\
                  + \rho\nabla\Phi_{\rm gal} &=& 0
\label{momentum}\\
\pop{\Eg}{t} + \diver{(\Eg{\bf u})}
          + \Pg\diver{\bf u} &=& \Gamma - \Lambda
\label{energy_g}\\
\pop{\Ec}{t} + \diver{\left((\Ec({\bf u}+\gc\bvA)\right)}
          + \Pc\diver{\bf u} & &  \nonumber \\
          - \bvA\nabla\Pc &=& \diver{(\bk\nabla\Ec)}
\label{energy_c}\\
\pop{\Ew}{t} + \diver{\left(\Ew({\bf u}+\bvA)\right)}
          + \Pw\diver{\bf u}  & &  \nonumber \\
          + \bvA\nabla\Pc &=& 0 \: .
\label{energy_w}
\end{eqnarray}
The Alfv\'en velocity is denoted by $\bvA$, the bulk gas velocity by $\vec{u}$, its density and pressure by $\rho$ and $\Pg$, respectively, while the CR and MHD wave pressures are labeled by $\Pc$ and $\Pw$. The galactic
gravitational potential is $\Phi_{\rm gal}$,  which we take from \citet{DB91}, who used 
a Miyamoto-Nagai potential for the disk and bulge, 
as well as a spherical dark matter component. Additional heating
$\Gamma$ and cooling $\Lambda$ can be included and an energy sink
for cosmic rays is incorporated by $\bvA\nabla\Pc$, transferring
energy from the cosmic rays to the waves due to the streaming
instability.
Energy sinks for the MHD waves other than adiabatic losses, e.g.\
ion-neutral or nonlinear Landau damping are not included
\citep[for a detailed discussion see][]{Ptus:97}. The former is
neglected because the level of ionization can be easily maintained
in a tenuous hot outflow plasma, and the latter is expected to definitely
become  important, once the fluctuation amplitude
reaches ${\delta \vec{B}/\vec{B}} \sim 1$, when quasilinear theory breaks down. 
Since we start with a negligible
wave field, assuming that all \emph{outgoing} waves are generated
by the resonant instability, damping conditions in most model runs occur only farther
out in the flow, where for example particle acceleration has
already taken place. However, in general the effect of nonlinear Landau damping should not be neglected 
\citep[see also discussion in][]{Ev:08}. Finally, in the case of
thermally driven winds, such as in starburst galaxies, the
contribution to the thermal pressure by wave damping is generally
small, because the fraction of CR energy density is lower than
the thermal one. 

The above system is closed by the equation of states relating the
pressures to the energies by
\begin{equation}  \label{e.eos}
 P_{\rm c,g,w} = (\gamma_{\rm c,g,w}-1)\, E_{\rm c,g,w} \: .
\label{close}
\end{equation}
Assuming ultrarelativistic particles, we adopt $\gc=4/3$, and
taking a monoatomic gas, we set $\gt=5/3$. A value of $\gw=3/2$
corresponds to a wave energy density $\Ew = \dBs /4\pi$ of the
mean squared fluctuating magnetic field component. Since the
magnetic wave pressure is given by $\Pw = \dBs /8\pi$, we obtaina value of 
$\gw=3/2$ by analogy. We also have to specify
the mean diffusion coefficient $\bk$ for the cosmic rays 
\citep{Dru}, which is averaged over the particle distribution
function (in space and momentum), and is usually taken to be a
free parameter. According to cosmic ray propagation models
\citep[e.g.,][]{Gin:Pu}, $\bk$ is estimated as an
order of $10^{29}\,{\rm cm}^2 {\rm s}^{-1}$. 

In the self-consistent 
CR propagation model of Ptuskin et al.~(1997), the field parallel diffusion 
coefficient is calculated from the MHD turbulence level arising 
from CR streaming, with generation balanced by 
nonlinear Landau damping, so that $\bk_{\parallel} \simeq 10^{27}  
\left(\frac{p}{Z m c}\right) ^{q-3} \, {\rm cm^2 s^{-1}}$, 
for Galactic halo conditions, 
where $q \approx 4$ is the power-law index in the particle momentum 
distribution near CR sources, and $p, m,$ and $Z$ are the particle momentum,
mass and nuclear charge, respectively.   
The influence of particle diffusion and acceleration is discussed in 
Sect.~\ref{s.diff} and more detailed properties of the system of
Eqs.~(\ref{mass})-(\ref{close}) can be found in \citet{DB91}.

These equations are solved in a 1D flux tube
geometry, where all quantities depend on the projected distance
$z$ from the galactic plane and the time $t$. To fix ideas we
specify the geometric properties of the flux tube, which
represents a transition from planar to spherical flow, and is in
its simplest form written as
\begin{equation} \label{e.tube}
A(z) = A_0\left[1 + \left(\frac{z}{z_0}\right)^2\right] \,,
\label{flux_tube}
\end{equation}
where $z_0$ defines a typical scale above which the flux tube
opens due to spherical divergence (typically of the order of a
galactic radius). For short distances $z\ll z_0$, the flow is
essentially planar and as seen e.g.~in Table~\ref{t.vel_mod} the lower values
of $z_0$ lead to faster shocks propagating along the flux tube. 
$A_0$ denotes the cross section at
our reference level of $z=0$, which cancels out in all equations. 

It has been emphasized by some authors that the equations
should be written and solved in general ellipsoidal
\citep{Ficht:91} or at least in oblate spheroidal coordinates
\citep{Ko}. A test of the simple formulation of Eq.~(\ref{flux_tube})
in comparison to oblate spheroidal coordinates has shown that the
maximum deviation is an enhancement of the mass loss rate by about
15\% for a flux tube at a Galactocentric distance of $R_0=10$ kpc 
\citep[]{Bre:94}.
The discrepancy even decreases for shorter distances, because the effect
of a radial (as opposed to vertical) flow component becomes progressively
weaker. In addition, it is more likely that the injection of supernova
heated gas into the base of the halo also imparts a $z$-component of momentum
to the flow, again enhancing the vertical component. Thus, for practical purposes, 
Eq.~(\ref{flux_tube}) seems fairly robust and time saving, and only a
complete solution of the axisymmetric MHD equations, in which the surfaces of magnetic
pressure are calculated self-consistently by a Grad-Shafranov type equation,
marks a substantial improvement. Here we are primarily concerned with
the interplay of various physical processes and the r\^ole of cosmic rays
as a driving agent, and so we take advantage of the simple but flexible
flux tube formulation in Eq.~(\ref{flux_tube}).

As seen e.g. in Eq.~(\ref{e.bfeld}), the assumed flux tube geometry 
also determines the radial structure of the magnetic field. However, edge-on
observations of spiral galaxies like NGC~5775 reveal  
large scale magnetic fields of similar topology in their galactic halo \citep{Soida}.
Also recent 3D-numerical simulations of a cosmic ray driven dynamo 
\citep{Kulpa} show that random magnetic fields will evolve into
large-scale, ordered magnetic structures that reveal a quadrupole-like
configuration with respect to the galactic plane. Observationally, as well 
as theoretically, assuming a prescribed flux-tube geometry thus seems
appropriate for simplifying the time-dependent evolution of galactic winds
through this 1D-geometry.

To summarize the previous section we have shown how to set up a
system of equations that describes a galactic outflow driven by
the interplay of the gravitational potential and the interaction
of thermal gas, cosmic rays, and waves in an ad-hoc prescribed
geometry. Several outflow simulations are discussed in detail
in the following sections. The time-dependent effects of
dissipation, heating and cooling, as well as particle acceleration
lead to a number of features, which in most cases can be followed
until a time-independent, stationary solution is reached.

\subsection{Boundary conditions}
\label{s.bound}

Boundary conditions close to the disk have to be specified, allowing for  a transition
from a subsonic to a supersonic outflow. In steady state this is only
possible if the outward pressure tends to zero at infinity, as shown by
\citet{Par:58} for the solar wind. The terminal wind velocity
can be easily estimated from Bernoulli's equation.
At the inner boundary
located at a reference level, we have to fix, e.g., the gas density,
the gas, and wave pressure. The cosmic rays are determined by an
outward directed flux
%
\begin{equation} \label{e.flux}
 {\bf F}_{{\rm c},0} = \frac{\gc}{\gc-1}({\bf u}_0+{\bvA}_0)P_{{\rm c},0} -
     \left.\frac{\bk}{\gc-1}\frac{\partial\Pc}{\partial z}\right\vert_0  \: ,
\end{equation}
where the first part describes the advected particles through the
combined wave speed ${\bf u}_0 + {\bvA}_0$, and the second term
represents the diffusive flux across the inner boundary. Since
energetic particles always leave the galaxy within a so-called
residence time (about $\sim 10^7$ yrs for the Milky Way), we have
to ensure $F_{{\rm c},0}>0$. If we use a typical length scale $L$ to
approximate the diffusive term
in Eq.~(\ref{e.flux}) by $P_{{\rm c},0}/L$, we can calculate on which scale
the cosmic ray flux would vanish, i.e.,~${F}_{{\rm c},0}=0$ or
$L\sim \bk/\gc({\bf u}_0+\bvA)$. Even in the case of low  
$\bk=10^{27}\,{\rm cm^2 s^{-1}}$, we obtain $L>1\,{\rm Mpc}$, hence
$F_{{\rm c},0}>0$ for all realistic galactic parameters since 
the Alfv\'en velocity is $|{\bvA}_0|=6.9\,{\rm km/s}$
for our initial conditions.
The last quantity to be specified at
the inner boundary is the velocity ${\bf u}_0$, which in case of a
steady-state solution is determined by the conditions at the critical point, 
thereby fixing the mass loss rate.

At the outer boundary we impose simple outflow conditions; 
i.e., no gradients should exist at large distances from the galactic
plane, typically at $z=1000\,{\rm kpc}$. 
As studied in various computations the details of the outer
boundary do not change the overall behavior of the solutions.
In steady-state flows, the domain of influence of the inner
boundary conditions only encloses the subsonic flow region. The
location of the critical point, however, is not completely independent 
of the outer boundary conditions. In particular, it is required that the 
pressure at the sonic point exceeds the pressure at the outer boundary. 
Otherwise, no transsonic flows are possible and only breeze solutions will
exist. For time-dependent flows such restrictions do not apply in
general, and only in the time-asymptotic limit do comparisons with
steady-state flows make sense.

As described in the following sections, we have computed
time-dependent galactic winds arising from changes of the inner 
boundary conditions. In particular, we studied how these variations
can be traced in the outflow and how the perturbed solution evolves towards
a stationary flow.

\subsection{Numerical method}
\label{s.num}

Since the numerical method has been described in detail 
\citep{Saas}, we emphasize here only some important features.
First, all equations are cast into a volume-integrated form to
ensure conservation of global quantities such as mass, momentum, and
energy. Such a finite volume formulation is also suited to an adaptive grid
that concentrates grid points to resolve steep gradients.
Second, the equations are discretized in an implicit way
so that for stationary solutions the time step exceeds the
typical flow time by several orders of magnitude. Third, an
artificial viscosity with a variable length scale in the tensor
formulation \citep{TW} is adopted to
broaden the shock waves over a finite number of grid points.
However, since we also use an adaptive grid, the typical broadening
length scale $l_{\rm visc}$ can be specified to be smaller than all physical
length scales, which enables correct computation of particle
acceleration; i.e., $l_{\rm visc} \ll \bk/\us $ with $\us$
denoting the shock speed. The first-order Fermi mechanism requires
that the energetic particles are exposed to a sharp discontinuity
to gain energy by being reflected between the upstream and
downstream media.

Applying the adaptive grid to these computations has the advantage
that the position of discontinuities can be located very precisely
through a clustering of grid points without specifying the type of
nonlinear wave beforehand. The propagation of
such nonlinear waves is plotted, e.g., in Fig.~\ref{f.rdisc_inn}, 
which enables a direct comparison with analytic estimates.

Computations based on an implicit method have the advantage of
allowing calculations of stationary solutions with the same numerical
method as for the time-dependent solutions. As a result several
solutions with fixed inner boundary values develop towards the
stationary solutions, all of which can be reached from different
evolutionary paths. We found from such a large number of calculations
that the stationary solutions look rather stable with respect to 
perturbations and that time-dependent flows relax to the
steady-state after the initial perturbations have propagated
through the computational domain. For our Milky Way with a
typical flow speed around $500\,{\rm km/s}$ we get a typical relaxation
time of about $10^9\,$years for a spatial region of $1\,{\rm Mpc}$.
The initial models can be taken directly from the time-independent solutions
of \citet{DB91}.

The actual discrete equations are then solved by a nonlinear 
Newton-Raphson procedure and 
are given in the appendices. The total number of spatial grid points
is set to $300$ throughout the computations. Due to our six physical
variables (gas density, gas velocity, thermal pressure, cosmic ray
pressure, wave pressure and radial position) at every grid point,  
each time step requires $1800$ new unknowns to be computed. 

\section{Results}
\label{s.res}

\subsection{Effects of cosmic ray diffusion}
\label{s.diff}

\begin{figure*}  
   \resizebox{\hsize}{!}{\includegraphics{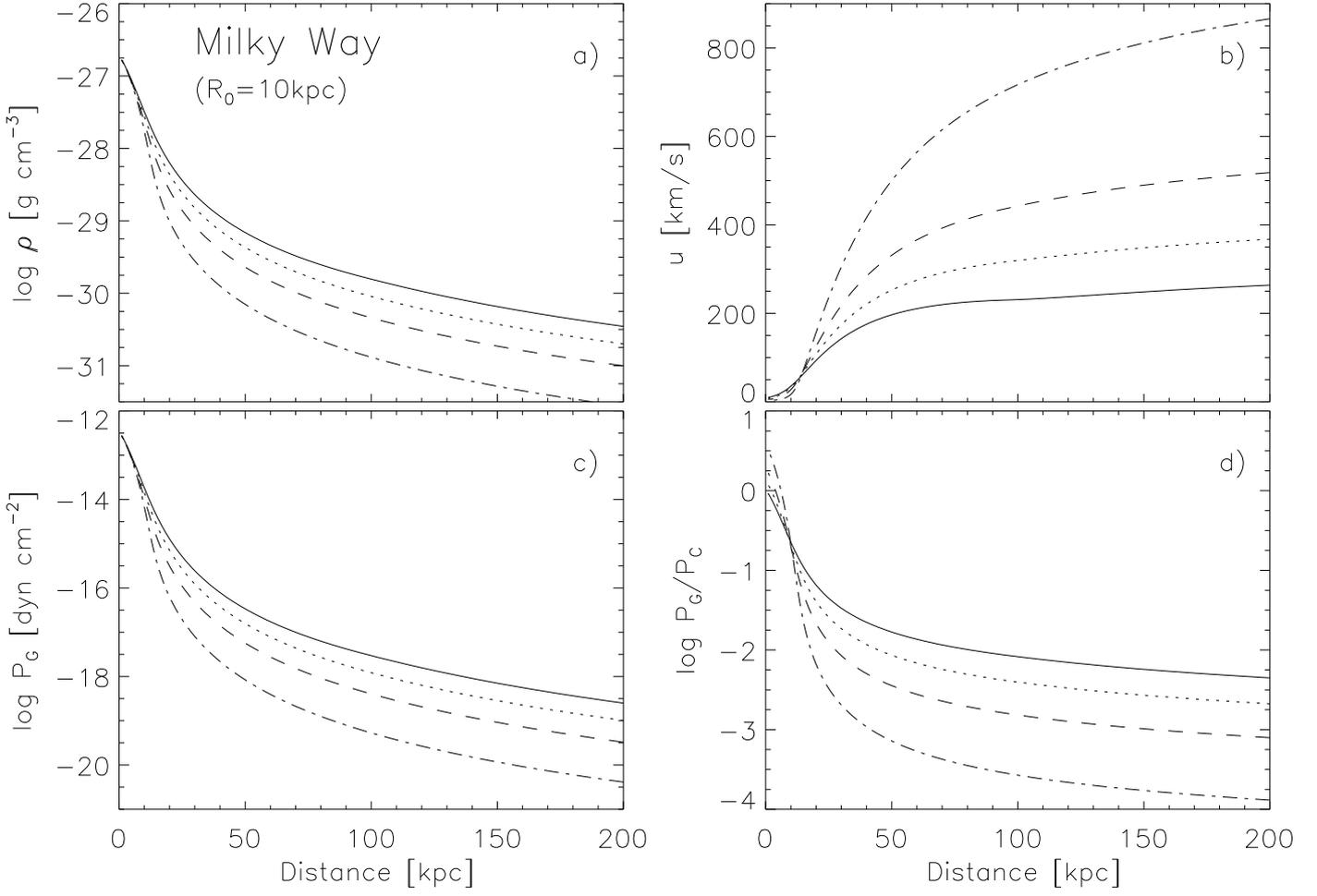}}
   \caption{The time-asymptotic flow structure of a galactic wind up to
            the innermost $200\,{\rm kpc}$ with different cosmic ray diffusion
            coefficients $\bk$.
            The case $\bk = 10^{27}\,{\rm cm^2\,s^{-1}}$ (full line) is almost
            identical to $\bk=0$, and noticeable differences occur only for 
            $\bk = 10^{29}\,{\rm cm^2\,s^{-1}}$ (dotted line),
            $\bk = 3\!\cdot\!10^{29}\,{\rm cm^2\,s^{-1}}$ (dashed line),
            and $\bk = 10^{30}\,{\rm cm^2\,s^{-1}}$ (dashed-dotted line). 
            The different panels represent:
            a)~gas density, b)~gas velocity, c)~thermal gas pressure,
            d)~ratio of gas pressure to cosmic ray pressure.}
   \label{f.diff}
\end{figure*}

Before we compute time-dependent solutions of the system of CR-hydrodynamics
in a flux tube geometry (Eqs.~\ref{mass}-\ref{energy_w}) we investigate the
effects of cosmic ray diffusion in steady state solutions. The additional 
diffusive term written as
\begin{equation}
 \diver{(\bk\nabla\Pc)}
\end{equation}
increases the order of the system of equations and no simple and
straightforward analysis of critical points can be used to
characterize the solutions. As emphasized earlier the implicit
character of our numerical methods makes it easy to
implement such an additional transport mechanism and in
Fig.~\ref{f.diff} the influence of a mean diffusion coefficient
$\bk$ is shown. 

The spatial wind profiles plotted in Fig.~\ref{f.diff} 
exhibit the changes of the overall flow structure caused
by different cosmic ray diffusion
coefficients in the astrophysical relevant parameter range of $\bk
= 10^{27} - 10^{30}\,{\rm cm^2 s^{-1}}$ within the innermost
$200\,{\rm kpc}$. Starting at a reference level of $1\,{\rm kpc}$,
we follow the flow up to a distance of $1000\,{\rm kpc}$ in order
to reach asymptotic values. For these computations the numerical
parameters correspond to our Galaxy, and we keep fixed most
quantities at the inner boundary, i.e.~the gas density of $\rho_0
= 1.67\!\cdot\!10^{-27}{\rm g\,cm^{-3}}$, the thermal gas pressure
$P_{{\rm g},0} = 2.8\!\cdot\!10^{-13}{\rm dyne\,cm^{-2}}$, the
vertical magnetic field component $B_0 = 1\,\mu{\rm G}$, and the
initial wave pressure $P_{{\rm w},0} = 4\!\cdot\!10^{-16}{\rm
dyne\,cm^{-2}}$, which corresponds to magnetic fluctuations at a
low level of $0.1\,B_0$.
In the case of no diffusion ($\bk=0$), the mass flux and therefore the
initial velocity $u_0$ are determined by the critical point (CP) conditions and
left as a free inflow boundary for the diffusive cases.
To investigate the effect of cosmic ray diffusion we fixed the
total cosmic ray flux $F_{{\rm c},0}$ at the inner boundary
through condition (\ref{e.flux}).

\begin{figure}   
  \resizebox{\hsize}{!}{\includegraphics{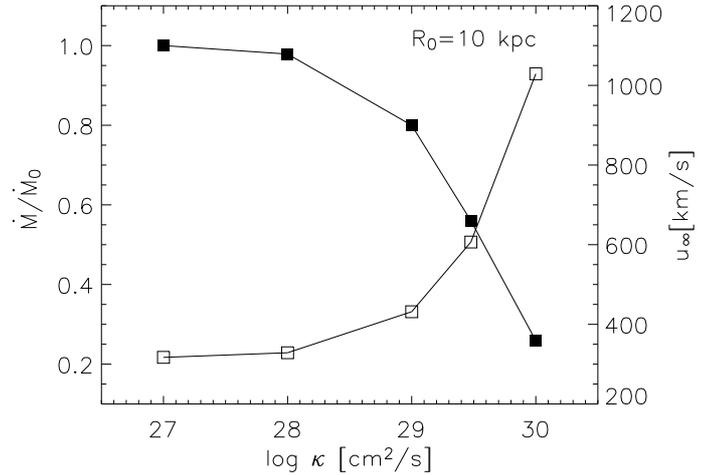}}
  \caption{Relative mass-loss rates (filled symbols) $\dot M$ in units
  of $\dot M_0$, the mass-loss rate at $\bk=0$, and corresponding terminal velocities
  (open symbols) in $[{\rm km/s}]$ for a flux tube located in our
  Galaxy at a galactocentric distance of $R_0=10\,{\rm kpc}$ 
  as a function of the mean cosmic ray
  diffusion coefficient $\bk$. Only values of 
  $\bk \ge 10^{29}{\rm cm^2\,s^{-1}}$ reduce the mass loss
  rate significantly, but also increase the final wind velocities owing to lower
  thermal gas ejections (see text for details).}
  \label{f.loss_cr}
\end{figure}

Since the diffusion length scale $L$ associated with a flow velocity
$U$ and diffusion coefficient $\bk$ is determined through $L\simeq
\bk/U$, i.e., for a typical value of $U\simeq 300\,{\rm km/s}$ and of 
$\bk\simeq 10^{29}\,{\rm cm^{2}s^{-1}}$, we get $L\simeq 1\,{\rm kpc}$, so 
significant modifications of
the large scale behavior are only expected for values of the
mean diffusion coefficient $\bk \gtrsim 10^{29}\,{\rm cm^{2}s^{-1}}$, as can be seen in
Fig.~\ref{f.diff} for the Milky Way. Since the
energetic particles can make an important contribution to the
overall galactic gas outflow, the coupling of CRs to the gas is an important process 
that alters the mass loss rate of the wind. 
In the case of diffusive losses, CRs can
escape from a galaxy without having to drag thermal gas along. 
To compare the solutions, the total cosmic ray flux 
$F_{{\rm c},0}$ is kept constant in Fig.~\ref{f.diff} 
and therefore an increase in the diffusive
losses has to be compensated by a corresponding decrease in the
advection of particles at the reference level; i.e., $u_0$
decreases. Since the mass flux (per unit area) is given by $\rho_0
u_0$, diffusion reduces the amount of thermal material within a
flux tube. Consequently, any increase in the diffusion coefficient
$\bk$ leads to lighter and faster winds with lower mass loss
rates.
Since the parameters for this computation are chosen for our own
galaxy similar to \citet{DB91}, cosmic rays
provide the dominant pressure contribution above $10\,{\rm kpc}$ for all values of $\bk$ as
can be inferred from Fig.~\ref{f.diff}d. This is because even for the highest value of $\bk$ 
the diffusion length does not exceed $10\,{\rm kpc}$. 
We emphasize that the solutions shown in Fig.~\ref{f.diff} for $\bk = 0$ for gas density,
outflow velocity, thermal, CR, and wave pressures, all as a function of 
distance, are in excellent agreement with the steady-state solutions 
obtained in \citet{DB91}.

The cosmic rays affects the inner boundary through the ingoing
cosmic ray flux $F_{{\rm c},0}$ (Eq.~\ref{e.flux}) as well as the transport of
material through the flux tube thanks to the available cosmic ray
pressure gradient. The increase in the mean cosmic ray diffusion
coefficient $\bk$ leads to changes in the mass-loss rate and the
final velocity as is shown in Fig.~\ref{f.loss_cr}. Since in
the adopted flux tube geometry for stationary winds the
mass-loss rate is given by
\begin{equation} \label{e.mloss}
 \dot M = A_0 \rho_0 u_0 = \pi R^2\rho_0 u_0 \: ,
\end{equation}
we have plotted in Fig.~\ref{f.loss_cr} the mass loss relative
to the case without cosmic ray diffusion, i.e.~$\bk=0$, and
denote this value by $\dot M_0$. Therefore we get rid of the
scaling by the initial cross section of the flux tube 
$A_0=\pi R^2$.  
This figure shows how
the total mass loss is decreasing as the relative contribution of
cosmic ray diffusion $\propto \bk\, \partial\Pc/\partial x$
increases compared to the advected flux $\propto (u_0 + \vA)P_{{\rm c},0}$ 
for a fixed value of the total CR-flux $F_{{\rm c},0}$ at the inner boundary 
(cf.~Eq.~\ref{e.flux}). 
As stated before, the typical length scale of cosmic ray interactions on a
flow with velocity $U$ is given by $\bk/U$, which alters the
cosmic ray pressure gradients available to drive the outflow. 
As seen in Fig.~\ref{f.loss_cr} values of $\bk \la 10^{29}{\,\rm cm^2\,s^{-1}}$ 
only have a small effect on the outflow characteristics, because 
such values are associated with length scales around $1\,{\rm kpc}$ 
taking a typical flow velocity of $U\sim 300\,{\rm km/s}$.
Comparing the results for various simulations of galactic winds from
our Galaxy (Fig.~\ref{f.diff}), we conclude that only mean cosmic ray
diffusion coefficients above $\bk \ga 10^{29}{\,\rm cm^2\,s^{-1}}$
affects the final outflow velocities and mass loss rates significantly.
Constraints based on galactic cosmic ray observations indicate a
diffusion coefficient on the order of $\sim 10^{29}\,{\rm cm^2\,s^{-1}}$, for
nucleons below about a GeV \citep[e.g.,][]{Axf:81}, 
which represent the bulk of the particles, and somewhat higher
values $\sim 10^{30}\,{\rm cm^2\,s^{-1}}$ for particles with higher
energies. 
In addition, the generation and propagation direction of
Alfv\'en waves are influenced by the cosmic ray gradients
\citep[e.g.,][]{DB91}. However, if parameters that are appropriate for different galaxies
only make a minor contribution to the wave pressure compared to the gas and cosmic
ray pressures, this effect is almost negligible here (see also
Figs.~\ref{f.rad_inn_struc}d and \ref{f.rad_struc}d). 

All results reported here were calculated for a Miyamoto-Nagai galactic potential
as used for our Milky Way at a galactocentric distance of 
$R_0=10\,{\rm kpc}$. 
This is slightly higher than the IAU value for the solar circle, but has been adopted here in order 
to compare results with \citet{DB91}. 
The dependence on the location within a galaxy enters only directly 
through a variation in the galactic potential $\Phi_{\rm gal}$, and indirectly by changes in gas 
density, pressure etc., as
discussed in \citet{DB91}. Integrating their galactic winds solutions
over the whole galactic disc yields 
a total mass loss rate of $\dot M_{\rm gal} \sim 0.3\,\Msol\,{\rm yr^{-1}}$.
This can serve as an upper limit of the \textit{global} galactic mass loss rate since the effect of
cosmic ray diffusion (Fig.~\ref{f.loss_cr}) will reduce this value, which was 
calculated from purely advective solutions.

Since the outflow velocity in case of thermally driven winds is usually more than an 
order of magnitude above the escape velocity, the mode of CR transport is 
immaterial here, since the mass loss rate itself is mainly determined by the 
thermal pressure. This is a consequence of the softer equation of state of the 
CRs, as compared to the thermal plasma, which leads to an efficient lifting 
of the halo weight at \textit{larger} heights, when 
the mass density is already smaller. In terms of galactic wind flows, the 
thermal plasma gradient lifts up the gas close to the disk, and the CRs act 
as a kind of afterburner, accelerating the sluggish thermal winds to higher 
velocities. This effect is obviously exacerbated by an increasing CR diffusion 
coefficient.
    
The flow time through the wind is given in our computational domain by
\begin{equation}    \label{e.flow_time}
  t_{\rm flow} = \int \frac{dz}{u(z)}  \;,
\end{equation}
where the boundaries have to be taken e.g. between $1\,{\rm kpc}$ and $1\,{\rm Mpc}$. 
In accordance with higher outflow velocities for increasing cosmic ray diffusive
transport the flow times up to $1\,$Mpc decrease from $4.1\cdot 10^{9}\,$years ($\bk=0$) to 
$2.8\cdot 10^{9}\,$years in the case of $\bk=10^{30}{\rm cm^2\,s^{-1}}$. The
flow times or final velocities do not scale directly with the mass loss rates (Fig.~\ref{f.loss_cr}) since 
increasing the diffusion $\bk$ changes the inner boundary conditions (Eq.~\ref{e.flux}) and 
also lowers the cosmic ray gradients needed to further accelerate the outflow. The
sonic point of the galactic winds shrinks from $108\,$kpc to $48\,$kpc.

\subsection{Time-dependent flow features}
\label{s.time}

\begin{figure*}   
   \resizebox{\hsize}{!}{\includegraphics{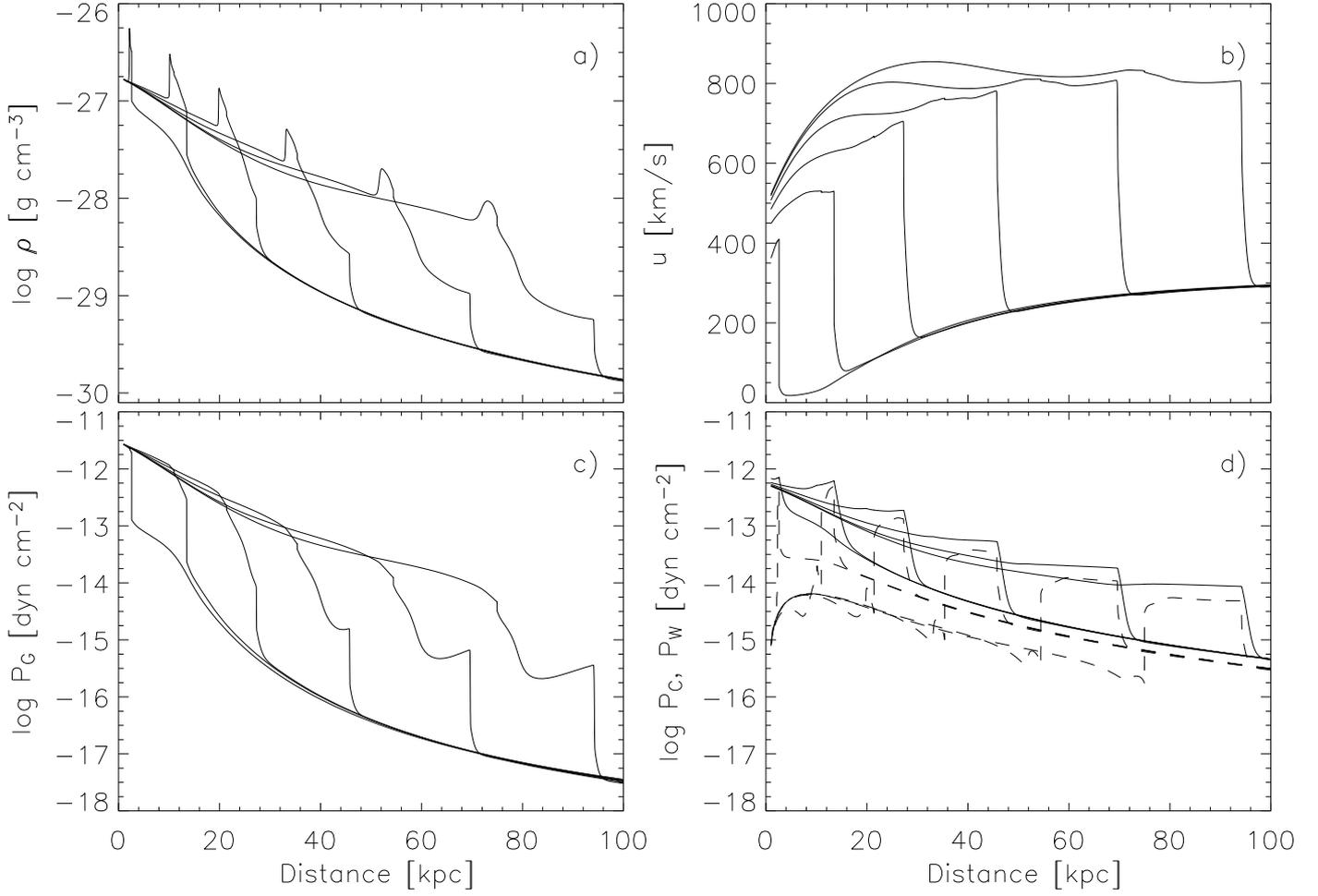}}
   \caption{The time-dependent flow structure of a galactic wind up to
            $100\,{\rm kpc}$ at six different times at
            $3.4\!\cdot\!10^6$, $2.0\!\cdot\!10^7$, $3.6\!\cdot\!10^7$, $5.4\!\cdot\!10^7$,
            $7.7\!\cdot\!10^7$ and $1.0\!\cdot\!10^8\,$years.
            At $t=0$ the gas and cosmic
            ray pressure has been increased by a factor of $10$ at the inner
            boundary of the flux tube, simulating a violent SN-explosion. Panel~a) 
            plots the development of the density structure where the forward as
            well as the reverse shock become visible. The velocity in $[{\rm km/s}]$
            is depicted in panel b) where the particle acceleration broadens the
            shock fronts in time.
            In panels c) and d) the curves correspond to the gas and the cosmic ray pressure (solid line)
            together with the wave pressure (dashed lines),
            respectively. The velocities
            of these features are also plotted in Figs.~\ref{f.rdisc_inn} and \ref{f.vel}.
            (see text for more details).}
   \label{f.rad_inn_struc}
\end{figure*}

Assuming that the flux tube geometry is given and fixed in time we
expect at least time-dependent inner boundary conditions. In a realistic case of repeated
SN-explosions or a starburst event (on larger scales), the inner variations will be
determined by the temporal history of such explosive events. Every change at the
base of the flow introduces additional disturbances, which then propagate outwards
and modify the flow conditions along the flux tube. To keep these structural changes simple and 
to illustrate the time-dependent behavior of a galactic wind, we have
chosen a stationary model as our initial model with a flux tube opening 
distance $z_0=15\,$kpc and a mean cosmic ray diffusion coefficient of 
$\bk=10^{29}\,{\rm cm^2/s}$, plotted also as dotted
line in Fig.~\ref{f.diff}. The propagation of waves is driven by a sudden increase
in gas and cosmic ray pressure at the inner boundary located at $1\,$kpc, where we
have raised the pressures by a factor of $10$ over a time interval of $10^6\,$years. All
other parameters are kept constant. Since we use an implicit numerical method, we can follow
the propagation of these pressure waves through the whole computational domain until a 
new stationary solution has been established within  
$1\le z/[{\rm kpc}]\le 1000$. Typically, the flow time (Eq.~\ref{e.flow_time}) up to $1\,$Mpc is
of a few $10^9\,$years.

As plotted in Fig.~\ref{f.rad_inn_struc} for the innermost $100\,{\rm kpc}$ 
the changes at the inner boundary result in a 
propagation of strong nonlinear waves along the flux tube since the conditions at the
base of a galactic wind have been altered dramatically. In the simulations reported here, we keep the
situation as simple as possible and see how pressure waves are initiated and generate
a flow pattern similar to a supernova remnant (SNR) \citep[e.g.,][]{Dor:91}. 
A strong forward shock wave propagates
through the unperturbed wind structure, accompanied by a reverse or
terminal shock, which slows down the newly developed outflow. However, owing to thehigh density
and pressure gradients within the initial flux tube, the overall properties of the
shock waves cannot be characterized by a rather simple self-similar behavior, which is
typical of the SNR case. 
At both shock waves, particles are accelerated, which modifies the shock 
structure itself, and the dissipation
of waves heats the thermal gas. The maximum momentum $\pmax$ of the accelerated particles 
attainable in such a shock configuration
is plotted in Fig.~\ref{f.pmax}, resulting in a reacceleration of the Galactic cosmic rays already
driving the outflow. 
At the moment we have not included any radiative losses of the
thermal plasma in our computations. 
Radiative cooling should not be dramatic close to the disk where the temperature 
is still above $10^6$ K and where the bulk of the particles are post-accelerated. 
At greater distances, when the temperature and density have dropped, cooling times 
(which are proportional to $1/\rho$) still remain high. 

\begin{figure}  
  \resizebox{\hsize}{!}{\includegraphics{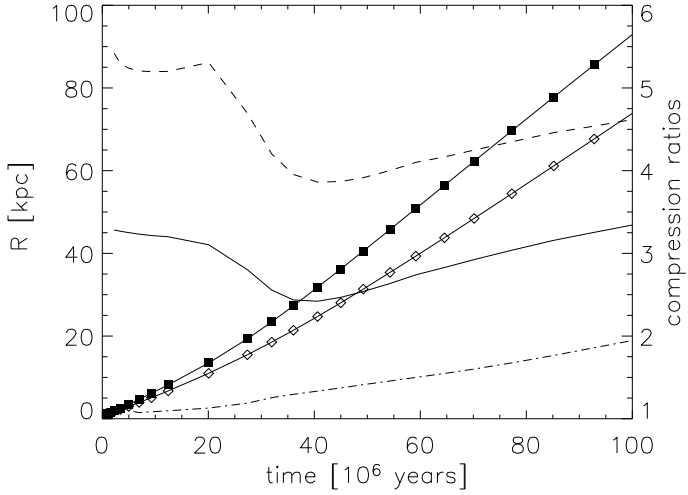}}
  \caption{The locations of the two shock fronts (forward shock: filled squares; 
  reverse shock: open diamonds) for galactic distances $z\le 100\,$kpc
  and compression ratios of the gas as a function of time.
  Already at early times ($t \le 20\!\cdot\! 10^6{\rm years}$) the shock
  propagation almost becomes constant. The solid line denotes the compression ratio of
  the forward gas subshock, the dashed line depicts the total compression
  ratio including the contribution form the cosmic ray precursor. The
  dashed-dotted line plots the gas compression ratio of the reverse shock 
  growing in time (see text for more details).} 
\label{f.rdisc_inn}
\end{figure}

In Fig.~\ref{f.rad_inn_struc}a the gas density clearly exhibits the separation of the flow
structures between $3.6\!\cdot\!10^7\,$years and $1.0\!\cdot\!10^8\,$years since the
forward shock propagates finally at a speed of $930\,{\rm km/s}$, whereas the reverse
shock stays behind at $840\,{\rm km/s}$. Already after $10^7\,$years the velocity difference
has accumulated to a spatial separation of about $1\,{\rm kpc}$.  The variation
in these velocities caused by different flux tube opening scales $z_0$ (see Eq.~\ref{flux_tube}) is
given in Table~\ref{t.vel_mod}. The gas compression ratio at the
forward shock varies in time (cf.~Fig.~\ref{f.rdisc_inn}) since the flow is 
strongly affected by the cosmic ray 
pressure contributions having an adiabatic index of $\gc=4/3$ as well as by the acceleration of
cosmic rays, which tends to broaden the shock transition (cf.~Fig.~\ref{f.rdisc_inn} 
for more details).

The gas velocity portrayed in panel~b) of Fig.~\ref{f.rad_inn_struc} exhibits the importance
of the forward shock where the outflow speed within the innermost $100\,$kpc 
increases from values less than $200\,{\rm km/s}$ to more than $800\,{\rm km/s}$. Although
visible in the gas density and gas pressure, the reverse shock plays a minor r\^ole for the
thermal gas during this initial expansion since the flows quickly accelerates down the 
large gradients. 
Only at distances $z\ga 200\,{\rm kpc}$ does the Mach number of the reverse shock increase 
again when the flow propagates in background with small gradients. The spatial variations in the
post-shock region are caused by gas pressure gradients (s. Fig.~\ref{f.rad_inn_struc}c), 
which are not smoothed by diffusive transport processes, such as the corresponding cosmic rays in
Fig.~\ref{f.rad_inn_struc}d, but are affected by additional heating from the dissipation of waves.
Due to the pressure increase at the base
of the flux tube the gas velocity at the inner boundary changes from 
$u_0=11\,{\rm km/s}$ to a new asymptotic value of $517\,{\rm km/s}$. 
Since the density $\rho_0$ is kept constant,  the mass loss
$\dot M \propto \rho_0 u_0$ also increases by a factor $517/11\sim 50$ from this flux tube 
into the intergalactic space.  Clearly, all the above values depend on 
the galactic potential, as well as on the flux tube opening $z_0$ (see Table~\ref{t.vel_mod}). 
As a result, these simulations may serve as a typical example for the case of our Galaxy.

The features in the gas pressure are dominated by the forward shock as shown in 
Fig.~\ref{f.rad_inn_struc}c within the innermost $100\,{\rm kpc}$. 
Since the flow is accelerating down the strong density gradient, the Mach number of the forward shock 
increases. As mentioned before, the reverse shock plays a minute r\^ole for the thermal gas at these stages 
as seen, e.g., in the rightmost curve by a small pressure jump at $t=4.8\!\cdot\!10^7\,$years, 
at a distance of $62\,{\rm kpc}$.

The cosmic ray pressure (Fig.~\ref{f.rad_inn_struc}d) dominates the flow 
structure as can be easily inferred by comparison with the gas pressure of panel~c). 
Energetic particles contributing to the cosmic ray pressure are mainly 
accelerated at the forward shock, and the shock structure is 
smoothed when the precursor region develops. Since we have adopted a 
mean cosmic ray diffusion coefficient of $\bk=10^{29}\,{\rm cm^2/s}$ 
the typical acceleration time scale yields $
\tau \sim \bk/\us^2\sim 1.5\!\cdot\!10^6\,$years for an 
initial shock velocity of $450\,{\rm km/s}$. 
The spatial scale corresponding to this acceleration
time is given by $l\sim \bk/\us\sim 0.8\,{\rm kpc}$, which is very close to 
the inner boundary. As also discussed in Sect.~\ref{s.acc} and depicted 
in Fig.~\ref{f.pmax} the particles are accelerated rapidly 
and close to the galactic plane. 
Although the reverse shock does not play an important r\^ole for the
overall cosmic ray pressure at the early outflow stages 
(see also Fig.~\ref{f.rdisc_inn})
a small number of individual particles 
will gain additional energy from this
second shock wave (see also Fig.\ref{f.pmax}).

In Fig.~\ref{f.rad_inn_struc}d we have also plotted the radial variations in the
wave pressure $\Pw$ by dashed lines, where the corresponding wave energy 
equation (\ref{energy_w}) contains
the dissipation term $\vA\nabla\Pc$ heating the thermal gas. The typical shape of these
curves clearly exhibits the location of the forward and the reverse shocks, seen as pronounced
changes of $\Pw$. We can also infer the importance of particle acceleration by the smoothed
edges. Again, the wind parameters typical of our Milky way show the importance of
the excited MHD-fluctuations $\Pw = \dBs /8\pi$ with $\Pw = (\gw-1)\Ew$ and $\gw=3/2$ for 
driving the outflow. Without the additional diffusive transport due to $\diver{(\bk\nabla\Ec)}$ 
acting on the cosmic ray pressure $\Pc$ the wave pressure $\Pw$ is largely affected by the
reverse shock seen, e.g., at $z=73\,$kpc for the rightmost plotted time of 
$t=1.0\!\cdot\!10^8\,$years. Within the zone of the two shock waves the wave pressure $\Pw$
remains above the thermal pressure $\Pg$.
However, we do not want to stretch this argument any further, because some care must be 
applied to the importance of wave pressure. As mentioned before, 
nonlinear Landau damping will limit wave growth severely, 
resulting in additional thermal heating of the plasma.

\begin{figure*} 
   \resizebox{\hsize}{!}{\includegraphics{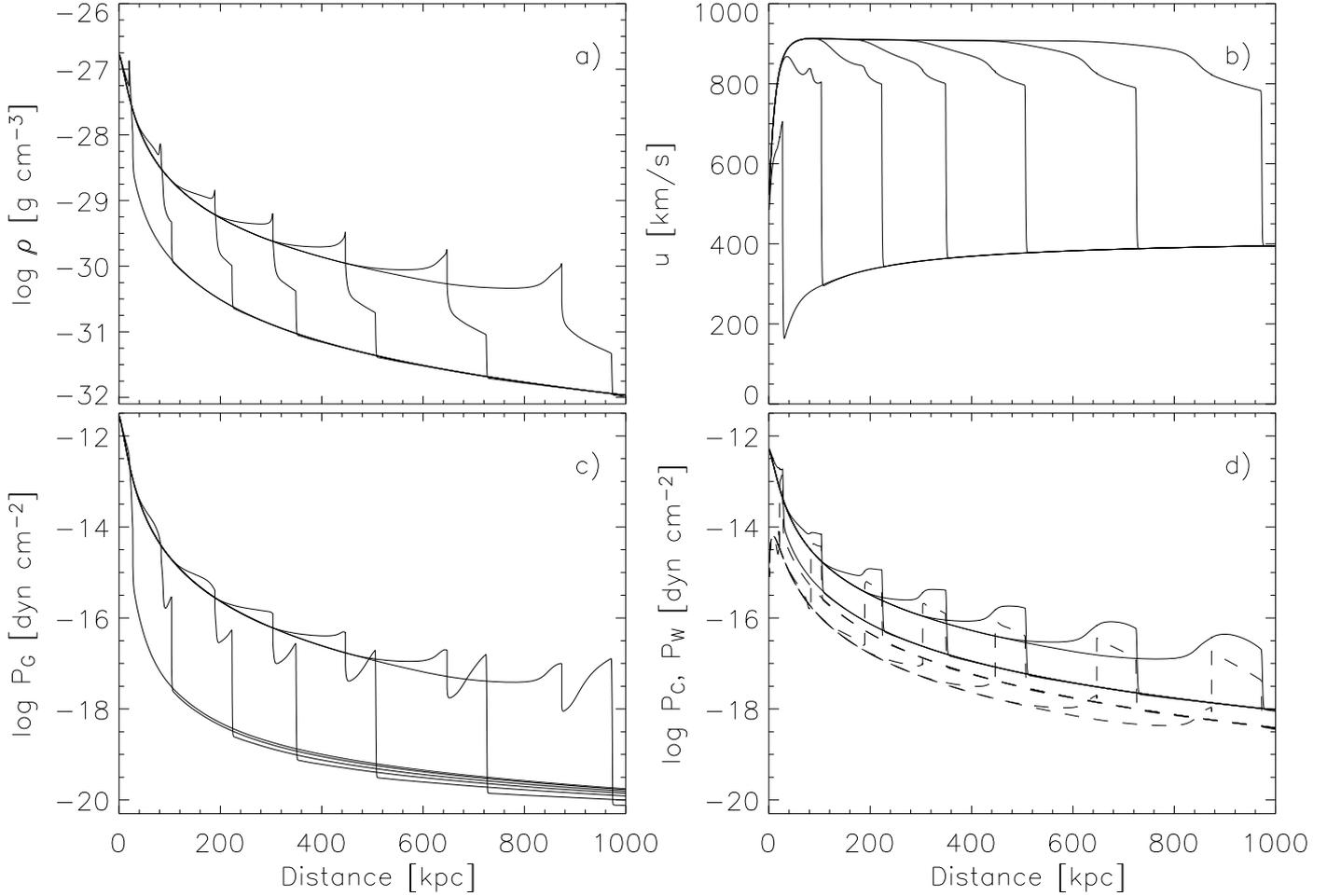}}
   \caption{The time-dependent flow structure of a galactic wind up to
            $1\,{\rm Mpc}$ at seven different times at 
            $3.7\!\cdot\!10^7$, $1.1\!\cdot\!10^8$,  $2.3\!\cdot\!10^8$,  $3.7\!\cdot\!10^8$,  
            $5.3\!\cdot\!10^8$,  $7.7\!\cdot\!10^8$, and $1.0\!\cdot\!10^9\,$years.
            At $t=0$ the gas and cosmic
            ray pressure has been increased by a factor of $10$ at the inner
            boundary of the flux tube, simulating a series of violent SN-explosions 
            within a short time interval. In Panel~a) 
            the development of the density structure is plotted, where both the forward and
            the reverse shocks become visible. The velocity in $[{\rm km/s}]$
            is depicted in panel b) where the particle acceleration broadens the
            shock fronts in time.
            In panels c) and d) the curves correspond to the gas and the cosmic ray pressure,
            together with the wave pressure (dashed lines),
            respectively. The velocities
            of the propagating features are also plotted in Figs.~\ref{f.rdisc_inn} and \ref{f.vel}
            (see text for more details).}
   \label{f.rad_struc}
\end{figure*}

Since the development of shock waves plays an important r\^ole in re-accelerating the
existing population of Galactic cosmic rays (see Sect.~\ref{s.acc}), we explored
the propagation of these flow features in Fig.~\ref{f.rdisc_inn} within the
first $10^8\,$years after our initial pressure increase 
at the inner boundary simulating multiple SN explosions. 
After the initial change, the inner boundary conditions are kept constant, 
causing nonlinear waves moving outwards at different speeds, leading to a 
growing separation between the flow features as can be 
seen in Fig.~\ref{f.rdisc_inn}. The velocities can
be directly estimated from the slopes and are given in the reference frame
of the galaxy and reach $930\,{\rm km/s}$ for the forward shock, and $840\,{\rm km/s}$
for the reverse shock, respectively. 
Unlike the solar or the galactic wind, 
where the termination shock is almost stationary with respect to the source frame, 
or in a supernova remnant, where the reverse shock is moving inwards thereby 
heating the ejecta, the reverse shock here is largely convected 
outwards by the galactic 
wind flow. In other words, its relative velocity with respect to 
the contact discontinuity ($\sim 80$ km/s) is much too low compared to the 
galactic wind speed of $\sim 900$ km/s, to propagate back to the disk. 
This is a direct consequence of the density profile in the undisturbed 
galactic wind medium, falling off as $1/z^2$ as we shall see. 
The shock velocities are constant, which can be easily understood in the 
framework of similarity solutions. If we disregard any processes containing 
explicit length and time scales during the formation of the flow, the propagation of a  
fluid element in an ambient medium 
with density decreasing by a power-law, $\rho \propto z^{\beta}$, is similar   
to a stellar wind type flow, in which the outer shock travels a distance 
$R_s \propto t^{\frac{3}{5+\beta}}$.
Since for an SN or starburst driven outflow, the terminal wind speed is 
reached very quickly, and therefore the steady-state wind density behaves as 
$\rho_w  \propto (u_w A(z))^{-1}$, where $A(z)$ is given by Eq.~\ref{flux_tube}. 
At some distance 
from the plane, $A(z) \propto z^{2}$, therefore $\beta = -2$, and thus  
$R_s \propto t$ or the shock velocity $u_s = dR_s/dt \sim const.$,
clearly visible after $t\ga 2\!\cdot\! 10^7\,$years or
$z\ga 12\,$kpc for our Milky Way parameter in Fig.~\ref{f.rdisc_inn} (filled squares).
The same must be true for the reverse shock (open squares in Fig.~\ref{f.rdisc_inn}), 
which sees a constantly decreasing density 
profile, again $\propto 1/z^2$, as it tries to propagate inwards, but is 
effectively convected outwards. This is also, by the way, the reason the 
reverse shock remains strong for a long time ($\sim 10^8\,$)years, 
because it does not run into medium with increasing 
density, where it would have to compress and set an increasing amount of 
mass into motion. 
However, the diffusive transport of cosmic rays, as well as the dissipative heating
of the thermal gas is important during the early expansion phases and the strength of
the reverse shock is therefore best visible in the wave pressure 
(Fig.~\ref{f.rad_inn_struc}d, dashed lines). 
This means that the reverse shock also remains a 
site of efficient particle 
acceleration, although particle escape would be primarily away from the galaxy 
similar to the galactic wind termination shock. 

The solid line in Fig.~\ref{f.rdisc_inn} represents the compression ratio of the
gas subshock within the forward shock (filled squares), 
also visible in Fig.~\ref{f.rad_inn_struc}a. 
The shock structure also exhibits a cosmic ray precursor where the
cosmic rays diffuse ahead of the wave thereby reducing the incoming plasma 
speed and converting adiabatically the kinetic energy into particle pressure. The
thermal gas is compressed in this precursor region, and the overall compression
ratio, including this effect, is plotted in the dashed curve 
of Fig.~\ref{f.rdisc_inn}, leading to compression ratios in excess of $4$. 
The variation in the compression ratios is due to 
the changing background of the initial flow structure (cf.~Fig.~\ref{f.diff}, dotted lines), 
which is basically confined to the innermost $50\,$kpc. The dashed-dotted
line depicts the gas compression ratio of the reverse shock, which increases
in time, since the wind velocity at the inner boundary rises, and the flow has to
be decelerated to the almost constant post shock velocity of the 
forward shock. We emphasize that both shock waves travel at different speeds, as
summarized in Table~\ref{t.vel_mod}, and that the most energetic particles 
can pick up the various velocity differences to get reaccelerated 
(see also Fig.~\ref{f.pmax}). 

In Fig.~\ref{f.rad_struc} the variables are plotted between the
inner boundary at $1\,{\rm kpc}$ and the outer boundary located
at $1000\,{\rm kpc}$ at seven different times. The first time step depicted corresponds to 
$3.6\!\cdot\!10^7\,$years, followed by $1.1\!\cdot\!10^8$, $2.3\!\cdot\!10^8$,
$3.7\!\cdot\!10^8$, $5.3\!\cdot\!10^8$, $7.7\!\cdot\!10^8$, and $1.03\!\cdot\!10^9\,$years.
The density, velocity, and
pressure structures in Fig.~\ref{f.rad_struc} clearly exhibit both
shock waves and the contact discontinuity, which is more difficult
to localize, since several pressure contributions are
important within this zone.  The slowdown
of the gas velocity of about $100\,{\rm km/s}$ is caused by the smoothed
reverse shock (Fig.~\ref{f.rad_struc}b). Both shock waves energize cosmic rays and 
in particular the further acceleration of particles at the reverse shock is clearly
visible through the local maxima in the corresponding $\Pw$-curves 
of Fig.~\ref{f.rad_struc}d. The increase in the particle pressure, as well as the
wave pressure (dashed lines) reveals that the reverse shock remains a site of efficient particle 
acceleration. Since after $t\ga 1.4\!\cdot\! 10^7\,$years, the reverse shock has reached a distance of
more than $100\,$kpc, the particles would tend to move away from the galaxy 
as at the galactic wind termination shock.

In the forward shock the
outflowing gas is pushed from $400\,{\rm km/s}$ to $800\,{\rm
km/s}$. Consequently, the overall flow time through the wind
up to $1\,{\rm Mpc}$ is shortened from the initial model of $3.2\!\cdot\! 10^9\,$years to
$1.08\!\cdot\! 10^9\,$years corresponding to a factor of $3$. The latter timespan is also 
identical to the time the initial perturbation needs to travel to the outer boundary located
at $1\,$Mpc.

\begin{figure}  
  \resizebox{\hsize}{!}{\includegraphics{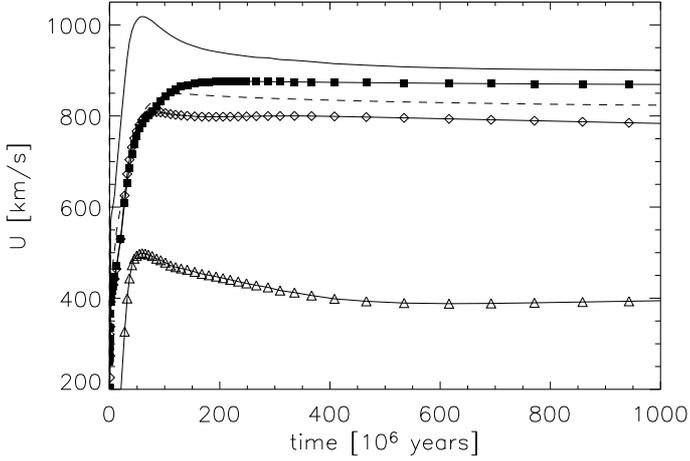}}
  \caption{The various velocities occurring in the time-dependent 
  solutions also plotted in Fig.~\ref{f.rad_struc}
  as a function of time. All velocities are given
  in the rest frame of the galaxy. The solid line depicts the velocity of the forward
  shock $v_{\rm s}$ and the dashed line the reverse shock $v_{\rm r}$.
  The highest gas velocity (filled squares) gives the upstream 
  velocity of the reverse shock $v_2$. The open triangles
  plot the post shock velocity $v_1$ of the reverse shock. 
  The open diamonds correspond to the gas velocity $v_0$ three diffusive 
  scales $\bk/\us$ in front of the forward shock.
  In less than about $7\cdot 10^7\,$years or $r < 50\,$kpc, 
  all the velocities become constant and are also given in
  Table~\ref{t.vel_mod}.}
  \label{f.vel}
\end{figure}

\section{Particle acceleration in galactic winds}
\label{s.acc}

As seen in the previous section, a temporal variation due to
changes in the star formation rate (and hence SN-rate) leads to an
increase in the gas pressure at the bottom of a flux tube. Such
time-dependent boundary conditions at the reference level result
almost inevitably in (several) shock waves propagating into the
halo (and possibly coalescing) as outlined in the previous section. Under such conditions 
(two shock waves separated by a contact
discontinuity, cf. Fig.~\ref{f.rad_struc}) energetic particles can
gain energy through the first-order Fermi-mechanism. 
The bulk of the particles, 
which enter the diffusive shock acceleration process, are already energized 
CRs from the disk sources so that there is no need to worry about injection. 
These particles instead undergo a reacceleration process. In contrast to the 
compression waves of the ``slipping interaction regions'' discussed in 
\citet{VZ:04}, the shock waves considered here can 
be strong. Since both forward and reverse shocks are running down a density 
gradient, they can strengthen considerably already close to the disk 
(see Fig.~\ref{f.rad_struc}). 

Diffusive particle transport along the magnetic
field lines of the flux tube in the time-dependent galactic wind flow 
thus opens the possibility
to accelerate cosmic rays to energies well beyond the knee, if large-scale
shock waves propagate through the wind. Adopting for simplicity the
test particle picture \citep[e.g.][]{Dru} we can integrate the 
equation
\begin{equation}    \label{e.pmax}
   \frac{d\pmax}{dt} = \frac{\pmax}{\tacc}
\end{equation}
to estimate the maximum momentum $\pmax$ reached during the
life time of a shock wave. We use the particle momentum in units of $mc$.
The diffusive acceleration time scale 
at a shock wave with upstream velocity $u_1$ and downstream
velocity $u_2$ for a planar shock (the radius of curvature is much larger than 
for supernova remnants) is given by
\begin{equation}  \label{e.tacc}
   \tacc = \frac{3}{u_1-u_2}
           \left(\frac{\kappa_1}{u_1}+\frac{\kappa_2}{u_2}\right) \: .
\end{equation}
The diffusion coefficients $\kappa_{1,2} = \kappa(p)$ depend on
the particle momentum $p$. If  fully developed turbulence is
assumed the so-called Bohm limit, where the mean free path is 
essentially reduced to a particle gyroradius, can be adopted, and for a
particle with mass $m$ and charge $Ze$ the diffusion coefficient
reads as
\begin{equation} \label{e.kapp}
   \kappa(p) = \frac{1}{3}\frac{mc^3}{ZeB}
               \frac{p^2}{\sqrt{1+p^2}} \: .
\end{equation}
The Bohm limit means that the field is completely random on a scale of the order 
of the gyroradius, and thus represents a minimum diffusion coefficient. 
It is indeed questionable whether higher energy particles find a wave field, 
which is random enough on larger scales for strong scattering to still be applicable 
\citep{Dog:etal:94}. On the other hand, in some models for accelerating CRs beyond the knee 
\citep{LB:00,BL:01}, it has been argued that while downstream of a shock wave, 
strong turbulence is excited and isotropization of the CR distribution function 
is rapid enough to promote recrossing, also nonlinear field amplification by CRs 
can occur upstream \citep[see also][]{MV:82}. 
Nonetheless, in our numerical calculations, we also have taken larger diffusion coefficients 
into account. As CR streaming generates waves for resonant interaction, it has been suggested
to assume that there is still a sufficient number of waves excited to guarantee pitch angle 
scattering, hence the validity of the first-order Fermi acceleration process.
As seen in the previous formula, the magnetic field strength $B$
enters through the diffusion coefficient $\kappa(p)$, and according
to $\diver{{\bf B}}= 0$, the adopted flux tube geometry
(Eq.~\ref{e.tube}) implies
\begin{equation}  \label{e.bfeld}
   A(z)B(z) = A_0B_0, \qquad B(z) =
   B_0\, \left[1+\left(\frac{z}{z_0}\right)^2 \right]^{-1} \: .
\end{equation}
As the flow and the shocks in a flux tube are propagating along the 
magnetic field, we are dealing with parallel shocks here 
with no field compression. It has been suggested though that the maximum energy 
gained by diffusive shock acceleration is higher for perpendicular shocks, when $\kappa$ is small, 
provided diffusion across the shock is fast enough to allow for multiple 
shock crossings \citep[e.g.][]{Jo:87}.

It is straightforward to understand the acceleration characteristics,  
if we take a look at sufficiently large time scales
$t \gg t_0$ and therefore $z\gg z_0$.  
As seen in Fig.~\ref{f.vel} the discontinuities are
propagating at constant velocities, so that 
we can write, e.g., the distance of a shock 
from the galactic plane as $z=\us t$, where $\us$ denotes the shock
velocity as given in Table~\ref{t.vel_mod}. As a result, Eq.~(\ref{e.kapp}) reduces via 
Eq.~(\ref{e.bfeld}) for large $z\gg z_0$ and $p\gg 1$ to 
\begin{equation} \label{e.kapp_simp}
   \kappa(p) \simeq \frac{1}{3}\frac{mc^3}{ZeB_0z_0^2}
              \,\us^2t^2p \: .
\end{equation}
Approximating also $\tacc \simeq 4\kappa(p)/\us^2$
we end up with a simplified equation for the maximum momentum $\pmax$
valid for the long term evolution
\begin{equation} \label{e.tacc_simp}
   \frac{d\pmax}{dt} = \frac{A}{t^2}, \qquad A = \frac{3ZeB_0z_0^2}{4mc^3}
                \: .
\end{equation}
This equation can be integrated easily, and specifying the initial conditions
by $p(t_0)=p_0$ at some time $t_0$, we get for $t\ge t_0$
\begin{equation} \label{e.pmax_sol}
   \pmax = p_0 + A\left(\frac{1}{t_0}-\frac{1}{t}\right)
               \: ,
\end{equation}
which leads to an almost constant maximum momentum $\pmax$ after the initial
acceleration.

\begin{figure} 
  \resizebox{\hsize}{!}{\includegraphics{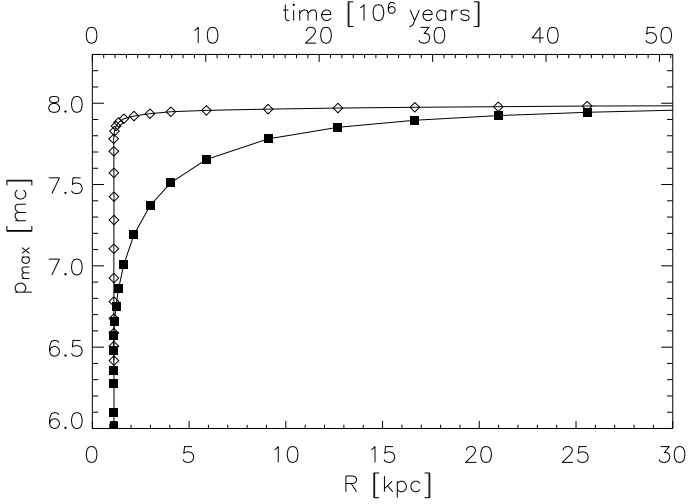}}
  \caption{The maximum momentum $\pmax$ in units of $[mc]$
           reached by a cosmic ray particle as a function of the
           shock distance $R_{\rm s}$ in $[{\rm kpc}]$ or time in $[10^6]\,$years.
           Already within a few kpc from the galactic
           plane (corresponding to a flow time of about 
           $10^7\,$years) an almost constant
           particle momentum is achieved. The squares are calculated
           due to particle acceleration occurring only at the forward shock,
           and the diamonds also include the acceleration at the reverse shock.}
  \label{f.pmax}
\end{figure}

This behavior of $\pmax$ is clearly visible in Fig.~\ref{f.pmax} where
Eq.~(\ref{e.pmax}) has been integrated in time for our velocity jumps. 
After the initial phase of a few $10^6\,{\rm years}$ the
momentum remains almost constant in accordance with the analytical 
estimate of Eq.~(\ref{e.pmax_sol}). As plotted in
Figs.~\ref{f.rad_inn_struc} and \ref{f.rad_struc}, the structure of the time-dependent flow
exhibits two shock waves, and energetic particles can be (re)accelerated at
both shock waves. The two
curves depict the differences between the cases of one and two shocks. 
The full squares show the maximum momentum of particles if only the acceleration at
the forward shock is considered; i.e., only the forward shock velocities $u_1$ and $u_2$ are
used in Eq.~(\ref{e.tacc}). The upper curve reaching even a 
higher particle momentum is computed by using the velocity difference between the
upstream region of the forward shock and the downstream region of the reverse shock. 
The acceleration at both shocks can therefore increase the maximum momentum by another 
factor of two compared to the acceleration only at the forward shock.
From Fig.~\ref{f.pmax} it is evident that this reacceleration 
of energetic particles is very rapid 
and occurs early on within a few kpc above the galactic plane. 

\begin{table*}
\caption{Asymptotic wave velocities in $[{\rm km/s}]$  in the frame of the Galaxy 
         and the maximal Mach numbers of the forward $M_{\rm f,max}$ and reverse shock 
         $M_{\rm r,max}$ for time-dependent galactic winds, and  
         the initial undisturbed wind velocity  $\uw$ at $1\,$Mpc. }
\label{t.vel_mod}
\centering
\begin{tabular}{c|c|c|c|c|c|c|c}
\hline
\hline
$z_0$  & $\uw$ & $v_1$ & $v_2$ & $\us$ & $u_{\rm r}$ & $M_{\rm f,max}$ & $M_{\rm r,max}$ \\
$[kpc]$& $[{\rm km/s}]$ &$[{\rm km/s}]$&$[{\rm km/s}]$&$[{\rm km/s}]$&$[{\rm km/s}]$&  & \\
\hline
  10 & 220 & 710 & 960 & 880 & 770 & 80 & 23 \\
\hline
  15 & 395 & 780 & 850 & 900 & 840 & 47 & 13 \\
\hline
  20 & 526 & 840 & 890 & 980 & 880 & 46 &  9 \\
\hline
\end{tabular}
\end{table*}

Since particle acceleration here is essentially a first-order Fermi process, 
we expect the usual power-law spectrum to emerge. It is known from simple 
one-dimensional diffusive shock acceleration theory for steady plane shocks 
that for particles injected at some momentum $p_0$, i.e. for a monoenergetic 
distribution function, a power-law distribution for the downstream spectrum, $f_2(p)$, 
will naturally arise, owing to the stochastic nature of the process, such that
\begin{equation}
f_2(p) = \frac{N_0}{4\pi}\frac{3 u_1}{u_1 - u_2}\left( \frac{p}{p_0}\right) ^\frac{3 u_1}
{u_1 - u_2} \,,
\label{p_spec} 
\end{equation}   
where  $N_0$ is the upstream particle number density with momentum $p \leq p_0$. 
The upstream and downstream velocities $u_{1,2}$ measured in the shock frame 
are given by 
\begin{eqnarray}
u_1 &=& \us  \left(1-\frac{\uw}{\us} \right) \\
u_2 &=& \frac{u_1}{r}  \,, 
\end{eqnarray} 
with $\uw$ the undisturbed wind speed, and $r$ the compression ratio.  
The power-law index in Eq.~(\ref{p_spec}) for a unmodified shock, i.e.~in one 
where the back-reaction of the particles on the shock structure can be still 
neglected, is then
\begin{equation}
q=\frac{3 u_1}{u_1 - u_2} = \frac{3( \gamma_g + 1) M^2}{2 (M^2 -1)} \,,
\end{equation} 
where the upstream Mach number is given by 
$M=\frac{\us}{\cs}\left(1-\frac{\uw}{\us}\right) \sim 0.5 \frac{\us}{\cs}$, 
for our standard case of $z_0=15\,$kpc (see also Table~\ref{t.vel_mod}); 
here $c_s$ is the local speed 
of sound ahead of the shock front. The galactic wind is already highly supersonic 
near the disk, but the Mach number depends on the relative velocity between the 
shock and the wind, and only if the flow, which is catching up, is much faster than 
the quiescent wind, can we expect a strong gas shock and $q \approx 4$ 
during the first 20 million years. 
However, the compression ratio of the forward shock is $r \sim 3$  (s.~Fig.~\ref{f.rdisc_inn}), 
implying $q=4.5$, which results in an energy spectral index of $\alpha \sim 2.9$, 
including an energy dependent diffusion $\propto E^{-0.65}$ of the particles propagating 
back to the disk. A mixture with CRs accelerated by the reverse shock 
with a lower compression ratio will tend to steepen the spectrum somewhat, 
which is already quite close to the observed value beyond the "knee".
Since the injected particles are those coming from the disk, and already having
a power-law spectrum, the resulting spectrum will also be a power-law. 
The spectral details will be the subject of further discussion in a forthcoming paper.

\section{Discussion and conclusions}
\label{s.con}

Soon after the diffusive shock acceleration theory was established as a major 
candidate to explain the origin of cosmic rays 
\citep{Kry, Axf:etal, Bel:a, Bel:b, Bla:Ost:78},
it became clear that supernova shock 
waves, which are the only non-exotic source of sufficient energy to provide 
the overall CR flux, would cut off in energy at about $10^{14}$ eV 
\citep{LC:83}, due to finite lifetime and curvature of the shock.
A more recent analysis shows that energies up to $Z \cdot 10^{15} eV$ 
are possible, if diffusion is at the Bohm limit \citep{Be:96}. 
The observed spectrum between the so-called knee and ankle thus have
remained unexplained for a long time. Recent X-ray observations 
of Tycho's SNR \citep{Eriksen} are interpreted such that particles are indeed
accelerated up to the ,,knee" in regions of enhanced magnetic turbulence
corresponding to particles energies in the range of $10^{14}\,$eV. However, 
the physical process of explaining energies of particles beyond the ``knee" up to the
``ankle" still remains unclear. We have already mentioned the problems 
of detecting particles accelerated at the galactic wind termination shock. 
As discussed previously, several interesting suggestions have been put forward recently 
to explain the spectrum between the ``ankle" and the ``knee".  
\citet{LB:00} picked up on an earlier idea \citep[e.g.,][]{Voe:mck}, 
arguing that the cosmic ray streaming itself would create sufficient turbulence upstream 
of the shock due to the rapid nonlinear growth of waves. 
If the SNR shock expanded into a stellar wind cavity, 
a maximum energy of $\sim Z \cdot 10^{18}\,$eV 
\citep{BL:01} could be achieved, beyond which CRs  
are commonly believed to be of extragalactic origin.

This hypothesis has received support in \citet{Dru:etal:03}. They actually 
calculated the resulting spectra from a Sedov-type shock in the framework of a simple 
``box'' model, in which the energetic particles are distributed uniformly within one 
diffusion length across the discontinuity, and are accelerated in momentum directly 
at the shock. They find a power-law spectrum with a small break at the knee all the way 
up to the ankle. The magnitude of the break, however, depends on the diffusion length and  
on scales up- and downstream of the shock. It is therefore not well determined in such 
a simple model. The break occurs at the correct energy (where ordinary SNR shocks cease
to accelerate), but the power-law exponent in the momentum distribution 
function can vary at the extreme between $-3.5$ (corresponding even to a hardening) and $-5.75$, 
while observations point to about $-4.5$.  Therefore more detailed and presumably 
numerical calculations are needed in order to quantify the implications of this model. 
Referring to the self-consistent Galactic CR propagation model by \citet{Ptus:etal:97}
\citet{VZ:04} have questioned the suggestion by \citet{BL:01}.
They argue that the transition boundary between diffusion and advection will have 
reached the termination shock at a distance of $\sim 300\,$kpc for particles with energies 
of  $\sim 10^{16} Z\,$eV and that particles with higher rigidities diffuse freely upstream 
of the termination shock and can only escape convectively downstream because of the 
small diffusion coefficient there, with no change in momentum. Thus the spectrum would 
be the same as the source spectrum in the disk, corresponding to a hardening of the 
$E^{-2.7}$-spectrum in the halo.  
In our model, acceleration occurs close to the disk, where the compression ratio is $\sim 3$,
making the spectrum naturally steeper near the sources than in the disk. 
In addition, the particles have to diffuse back to the disk, leading to further steepening of the spectrum, 
as discussed in the previous section. 
%
A detailed discussion of this problem will be presented  
in a sequel paper. Here we would just like to emphasize that time-dependent changes 
in the boundary conditions of the galactic wind, naturally lead to the formation of 
strong shocks that will reaccelerate particles to energies well above the knee. We  
agree with \citet{VZ:04} that the analysis of the cosmic ray spectrum 
has to include the propagation effects of particles not only in the disk, but also in the 
Galactic halo.  While we have analyzed these effects for starburst regions, it is 
clear that a similar scenario must hold for our Galaxy, although presumably restricted 
to smaller regions in the underlying Galactic disk. Nevertheless, 3D numerical 
simulations by \citet{AB:04} have shown that outflow occurs 
even for thick Lockman and Reynolds type gas disks. The crucial point is that 
once overpressured hot regions have punched a hole in the lower halo, material 
can travel almost unimpeded to considerable heights, resulting in an outflow 
once the cosmic ray pressure gradient takes over. These outflows will be modulated 
in a similar manner as the starburst types, due to changes in the star formation rate. 
The resulting shock waves can also be strong, because as we have emphasized, it 
is the \textit{relative velocity} between fast and slow winds that matters, and steepening 
of the shock wave will occur as it runs down the galactic wind density gradient.  

In addition to the dynamical phenomena caused by SN-explosions, the magnetic field
may also cause changes in the dynamics of the flow, and the idealized geometry will only 
approximate reality on average and on large scales. On smaller scales the field geometry can be quite 
complex, as shown by 3D numerical MHD simulations \citep[e.g.]{Ko:etal:99, AB:05}. 
The tangling of field lines will inevitably lead to a change in topology by magnetic reconnection \citep{Pa:92}
and to additional heating.  Also MHD wave heating of the halo has been considered to be responsible 
for highly ionized species \citep{Ha:83}. Field diffusivity and turbulence, which are a natural consequence 
of a SN-driven ISM \citep{AB:07}, will further promote dynamo action and 
B-field amplification \citep[e.g.]{Gr:etal:08}. 
Magnetized winds in other galaxies, e.g. dwarf galaxies, have also been modeled \citep{DT:10} 
in order to study the enrichment of the intergalactic medium with magnetic 
fields within a cosmological context.

%
%
Our model treats the base of the halo as an inner boundary condition, from which the flow emanates. In reality, there is a transition between disk and halo, and the outflow is a consequence of time-dependent SN activity in the disk. Again, a 3D numerical simulation would be necessary to properly model fountain flows \citep[e.g.]{AB:04}. 
However, none of the 3D models put forward so far has considered the effect of CRs and their nonlinear coupling to the flow, which is the purpose of this analysis.

Although also neglecting effects of Galactic rotation and nonlinear wave damping, our model 
has the advantage that particle acceleration is calculated self-consistently along with the 
galactic wind flow. The hydrodynamic version of the transport equation that we use 
even allows us to enter the nonlinear regime, where the shocks can be modified by 
the increasing CR pressure, so we are not restricted to test particle arguments. 
Such a self-consistent picture  allows us not only to deduce interesting quantities like 
mass-loss rate of a galaxy, but also to calculate e.g.\  X-ray spectra that can be compared 
to XMM-Newton and Chandra observations, in particular for delayed recombination from
nonequilibrium distributions of ions in the halo \citep[cf.]{Br:03}. In a next step 
we will include the CR momentum distribution function to numerically calculate the 
resulting cosmic ray spectra between the ``knee'' and the ``ankle''. \\

To summarize, in this paper we have shown that 
\begin{enumerate}
\item time-dependent galactic winds have an asymptotic behavior that matches 
 the steady-state solutions obtained by \citet{DB91} very well;

\item these steady-state solutions are stable with respect to perturbations;

\item changes in the inner boundary conditions are very likely to occur, since 
the flow time of the wind gas is on the order of $10^8$ years, so much longer 
than the switch-on time for superbubbles or starbursts;

\item changes in the boundary conditions, in particular an increase in pressure due 
to increased star formation rate, result in time-dependent features like shocks (forward 
and reverse) and contact discontinuities;

\item these shocks running down a density gradient can become very strong and 
propagate with roughly constant velocity;

\item cosmic rays, propagating along with outflowing gas from the disk, can be accelerated 
very efficiently to energies up to $10^{17} - 10^{18} Z\,$eV by these shocks, 
thus suggesting a new mechanism for generating the Galactic cosmic rays between the 
``knee'' and the ``ankle'';

\item particle acceleration is very fast and occurs close to the 
  galactic disk, typically with a few kpc; 

\item the particle spectrum at the shock is most likely steeper than in the disk, 
because the Mach number of the forward shock depends on the \textit{relative velocity} 
between the background wind and compressional waves, and is therefore smaller. 
This leads to a spectral index $ < -4$, i.e. a steepening at the shock. 
In particular, in the Milky Way, where bursts of star formation are less violent, 
shocks forming in the halo will have lower Mach numbers, thus leading to a steepening of 
a spectrum beyond the "knee";

\item in addition the spectrum must also steepen, because the particles 
have to diffuse back to the disk; thus, the lower galactic halo seems to be a 
promising site for accelerating CRs beyond the "knee".

\end{enumerate}

\begin{acknowledgements}

Part of this work has been supported by the Jubil\"aums\-fonds der
Oester\-rei\-chi\-schen National\-bank under pro\-ject
number~4605. The publication is supported by 
the Austrian Science Fund (FWF).
DB acknowledges support from the Bundesministerium
f\"ur Bildung und Forschung (BMBF) by the Deutsches Zentrum f\"ur
Luft- und Raumfahrt (DLR) under grant 50 OR 0207 and the
Max-Planck-Gesellschaft (MPG), where part of this work was already carried out.
\end{acknowledgements}

\bibliographystyle{apj} 
\bibliography{apj-jour,db1_ref}

\begin{thebibliography}{}
\bibitem[Adelberger \& Steidel(2000)]{AdSt:00}
    Adelberger K.~L., Steidel, C.C. 2000, \apj, 544, 218
\bibitem[Axford et al.(1977)]{Axf:etal}
    Axford W.I., Leer E., Skadron G. 1977,
    Proc. 15th Int. Cosmic Ray Conf. (Plodiv) 11, 132
\bibitem[Axford(1981)]{Axf:81}
    Axford W.I. 1981, Proc. Int. School and Workshop on Plasma
    Astrophysics, Varenna, ESA~SP--161, 425
\bibitem[Bell(1978a)]{Bel:a}
    Bell A.R. 1978a, \mnras, 182, 147
\bibitem[Bell(1978b)]{Bel:b}
    Bell A.R. 1978b,\mnras, 182, 443
\bibitem[Bell(1987)]{Bel}
    Bell A.R. 1987, \mnras, 215, 615
\bibitem[Bell \& Lucek(2001)]{BL:01}
    Bell A.R., Lucek, S.G. 2001, \mnras, 321, 433
\bibitem[Berezhko(1996)]{Be:96}  
    Berezhko, E. G. 1996, Astroparticle Phys. 5, 367
\bibitem[Berezhko et al.(1994)]{Ber:etal}
    Berezhko E.G., Yelshin V.K., Ksenofontov L.T. 1994,
    Astroparticle Phys. 2, 215
\bibitem[Blandford \& Eichler(1987)]{Bla:Eic}
    Blandford R.D. Eichler, D. 1987, Phys. Rep. 154, 1
\bibitem[Blandford(1988)]{Bla}
    Blandford R.D. 1988, in: Supernova Remnants and the Interstellar
    Medium. Roger R.S., Landecker T.L. (eds.) Cambridge Univ. Press,
    Cambridge, p. 309
\bibitem[Blandford \& Ostriker(1978)]{Bla:Ost:78}
    Blandford R.D., Ostriker, J.P. 1978, \apj, 221, L29
\bibitem[Bregman et al.(1992)]{Bre:etal}
    Bregman J.N., Hogg D.E., Roberts M.S. 1992, \apj, 387, 484    
\bibitem[Bregman(1980)]{Bre:80}
    Bregman J.N., 1980, \apj, 236, 577
\bibitem[Breitschwerdt(2003)]{Br:03} 
    Breitschwerdt D.  2003,  Rev. Mex. Astron. Astrophys., 15, 311    
\bibitem[Breitschwerdt(1994)]{Bre:94} 
    Breitschwerdt D., 1994,  Habilitation Thesis, University of Heidelberg, 158 p.
\bibitem[Breitschwerdt et al.(2002)]{Br:02}
    Breitschwerdt D., Dogiel, V.A., V{\" o}lk, H.J. 2002, \aap, 385, 216
\bibitem[Breitschwerdt et al.(1991)]{DB91}
    Breitschwerdt D., McKenzie J.F., V\"olk H.J. 1991, A\&A, 245, 79 
\bibitem[Breitschwerdt et al.(1993)]{DB93}
    Breitschwerdt D., McKenzie J.F., V\"olk H.J. 1993, A\&A, 269, 54
\bibitem[Breitschwerdt \& Schmutzler(1994)]{BS:94}
    Breitschwerdt D., Schmutzler M. 1994, \nat, 371, 774
\bibitem[Breitschwerdt \& Schmutzler(1999)]{BS:99}
    Breitschwerdt D., Schmutzler M. 1999, A\&A, 347, 650
\bibitem[Canizares et al.(1987)]{Can:etal}
    Canizares C.R., Fabbiano G., Trinchieri G. 1987, \apj, 312, 503
\bibitem[Dawson et al.(2002)]{Daw:02}
    Dawson S., Spinrad H., Stern D., Dey A., van Breugel W., de Vries W., Reuland M.
    2002, \apj, 570, 92
\bibitem[Avillez(2000)]{deA:00}
    de Avillez M. 2000, \apss~272, 22
\bibitem[Avillez \& Breitschwerdt(2004)]{AB:04}
    de Avillez M., Breitschwerdt D., 2004, A\&A, 425, 899  
\bibitem[Avillez \& Breitschwerdt(2005)]{AB:05}
    de Avillez M., Breitschwerdt D., 2005, A\&A, 436, 585  
\bibitem[Avillez \& Breitschwerdt(2005)]{AB:07}
    de Avillez M., Breitschwerdt D., 2007, \apjl, 665, 35  
\bibitem[Dogiel et al.(1994)]{Dog:etal:94}
    Dogiel, V.A., Gurevich, A.V., Zybin, K.P., 1994, A\&A, 281, 937
\bibitem[Dorfi(1990)]{Dor:90}
    Dorfi E.A. 1990, A\&A, 234, 419
\bibitem[Dorfi(1991)]{Dor:91}
    Dorfi E.A. 1991, A\&A, 251, 597
\bibitem[Dorfi(1998)]{Saas}
    Dorfi E.A. 1998, in: Computational Methods for Astrophysical Fluid Flows,
    27th Saas Fee Course, Springer, Berlin, p. 263
\bibitem[Dorfi \& Drury(1987)]{DD}
    Dorfi E.A., Drury L.O'C. 1987, J. Comp. Phys. 69, 175 
\bibitem[Drury(1983)]{Dru}
    Drury L.O'C. 1983, Rep. Prog. Phys. 46, 973   
\bibitem[Drury \& V"olk(1981)]{Dru:Voe}
    Drury L.O'C., V\"olk H.J. 1981, \apj, 248, 344
\bibitem[Drury et al.(2003)]{Dru:etal:03}
    Drury L.O'C, van der Swaluw, E., Carroll, O., 2003, astro-ph/0309820
\bibitem[Dubois \& Teyssier(2010)]{DT:10}
   Dubois, Y., Teyssier, R., 2010, A\&A, 523, 72
\bibitem[Eriksen et al.(2011)]{Eriksen}
    Eriksen A.K.,  Hughes J.P., Badenes C., Fesen R., Ghavamian P., 
    Moffett D., Plucinksy P.P., Rakowski C.E., Reynoso E.M., Slane P. 2011, \apjl, 728, 28
\bibitem[Everett et al.(2010)]{Ev:10}
    Everett, J.E., Schiller, Q.G., Zweibel, E.G. 2010, ApJ, 711, 13 
\bibitem[Everett et al.(2008)]{Ev:08}
    Everett, J.E., Zweibel, E.G., Benjamin, R.A. et al. 2008, \apj, 674, 285  
\bibitem[Fichtner et al.(1991)]{Ficht:91}
    Fichtner H., Neutsch W., Fahr H.J., Schlickeiser R. 1991, \apj, 371, 98
\bibitem[Ginzburg \& Ptuskin(1985)]{Gin:Pu}
    Ginzburg V.L., Ptuskin V.S. 1985, Sov.Sci.Rev.E. \apss, 4, 161
\bibitem[Gressel et al.(2008)]{Gr:etal:08}
    Gressel, O., Elstner, D., Ziegler, U. R\"udiger, G., A\&A, 486, L35
\bibitem[Habe \& Ikeuchi(1980)]{HI:80}
    Habe A., Ikeuchi S. 1980, Rep.Prog.Theor.Phys. 64, 1995
\bibitem[Hartquist(1983)]{Ha:83}
    Hartquist T. 1983, \mnras, 203, 117
\bibitem[Ipavich(1975)]{Ip:75}
    Ipavich, F. 1975, \apj, 196, 107 
\bibitem[Jokipii (1987)]{Jo:87}
    Jokipii J.R. 1987, \apj, 313, 842  
\bibitem[Jokipii \& Morfill(1985)]{JM:85}
    Jokipii J.R., Morfill G.E. 1985, \apj, 290, L1    
\bibitem[Jokipii \& Morfill(1987)]{JM:87}
    Jokipii J.R., Morfill G.E. 1987, \apj, 312, 170    
\bibitem[Kahn(1981)]{Kahn}
    Kahn F.D. 1981, in: Investigating the Universe, F.D. Kahn (ed.),
    D. Reidel Publ. Comp., Dordrecht, p. 1
\bibitem[Kapferer et al.(2006)]{Kap:06}
    Kapferer W., Ferrari, C., Domainko, W., Mair, M., Kronberger, T., Schindler, S., Kimenswenger, S., 
    van Kampen, E., Breitschwerdt, D., Ruffert, M. 2006, A\&A, 447, 827
\bibitem[Kikuchi et al.(1999)]{Kik:99}
    Kikuchi K., Furusho T., Ezawa H., Yamasaki N.Y., Ohashi T., Fukazawa Y., Ikebe Y. 1999, 
    \pasj, 51, 301
\bibitem[Ko(1991)]{Ko}
    Ko C.M. 1991, A\&A, 242, 85    
\bibitem[Korpi et al.(1999)]{Ko:etal:99}
    Korpi, M.J., Brandenburg, A., Shukurov, A., Tuominen, I., Nordlund, A., \apjl, 514, 99 
\bibitem[Krymsky(1977)]{Kry}
    Krymsky G.F. 1977, Dokl.~Nauk.~SSR~234, 1306,
    (Engl.~trans. Sov.~Phys.~Dokl.~23, 327)
\bibitem[Kulpa-Dybel et al.(2011)]{Kulpa}
    Kulpa-Dybell K., Otmianowska-Mazur K., Kulesza-{\.Z}ydzik B., Hanasz M., Kowal G.,
    W{\'o}lta{\'n}ski D., Kowalik K. 2011, \apj, 733, L18
\bibitem[Kulsrud \& Pearce(1969)]{Kul:Pea}
    Kulsrud R.M., Pearce W.D. 1969, \apj, 156, 445
\bibitem[Lagage \& Cesarsky(1983)]{LC:83}
    Lagage P.O, Cesarsky C.J. 1983, A\&A, 125, 249L
\bibitem[Lerche \& Schlickeiser(1982)]{LS:82}
    Lerche I., Schlickeiser R. 1982, \aap, 107, 148
\bibitem[LeVeque(1990)]{LeV}
    LeVeque, R.J. 1990, Numerical Methods for Conservation Laws,  
    Birkh\"auser-Verlag, Basel
\bibitem[Loewenstein(2001)]{Loew:01}
    Loewenstein, M. 2001, \apj, 557, 573
\bibitem[Lucek \& Bell(2000)]{LB:00}    
    Lucek S.G., Bell A.R. 2000, \mnras, 314, 65L
\bibitem[Markiewicz et al.(1990)]{Mar:etal}
    Markiewicz W.J., Drury L.O'C, V\"olk H.J. 1990, A\&A, 236, 487
\bibitem[McKee(1988)]{McK}
    McKee C.F. 1988, in: Supernova Remnants and the Interstellar
    Medium. Roger R.S., Landecker T.L. (eds.) Cambridge Univ. Press,
    Cambridge, p. 205 
\bibitem[McKee \& Cowie(1977)]{McK:Cow}
    McKee C.F., Cowie L. 1977, \apj, 215, 213
\bibitem[McKee \& Ostriker(1977)]{McK:Ost}
    McKee C.F., Ostriker J.P. 1977, \apj, 218, 148
\bibitem[McKenzie \& V\"olk(1982)]{MV:82}
    McKenzie, J.F., V\"olk, H.J. 1982, A\&A, 116, 191
\bibitem[Molendi et al.(1999)]{Mol:99}
    Molendi S., de Grandi S., Fusco-Femiano R., Colafrancesco S., Fiore F., Nesci R.,
    Tamburelli F. 1999, \apjl, 525, L73    
\bibitem[Normandeau et al.(1996)]{Nor:96}
    Normandeau M., Taylor A.R., Dewdney P.E. 1996, \nat, 380, 687 
\bibitem[Parker(1992)]{Pa:92}
   Parker, E.N., \apj, 401, 137
\bibitem[Parker(1958)]{Par:58}
    Parker, E.N. 1958, \apj, 128, 664 
\bibitem[Ponman et al.(1999)]{Pon:99}
    Ponman T.J., Cannon D.B., Navarro J.F. 1999, \nat, 397, 135
\bibitem[Ptuskin(1997)]{Ptus:97}
    Ptuskin V.S. 1997, Adv. in Space Res. 19, 697
\bibitem[Ptuskin et al.(1997)]{Ptus:etal:97} 
    Ptuskin V.S., V\"olk, H.J., Zirakashvili, V.N., Breitschwerdt, D., A\&A, 321, 434   
\bibitem[Reynolds et al.(2001)]{Rey:01}
    Reynolds R.J., Sterling N.C., Haffner L.M., 2001, \apjl, 558, L101
\bibitem[Sarazin \& White(1988)]{Sar:Whi}
    Sarazin C.L., White R.E. 1988, \apj, 331, 102
\bibitem[Soida et al.(2011)]{Soida}
    Soiada M., Krause M, Dettmar R.-J., Urbanik M. 2011, A\&A, 531, 127
\bibitem[Steidel et al.(2010)]{Steid:10}
    Steidel, C.C., Erb, D.K., Shapley, A.E., Pettini, M., Reddy, N., Bogosavljevic, M., Rudie, G.C., Rakic, O. 2010, ApJ 717, 289
    Steidel C.C., Giavalisco M., Pettini M., Dickinson M., Adelberger K.L. 1996, \apjl, 462, L17
(Steidel et al. 2010, ApJ 717, 289).
\bibitem[Steidel et al.(1996)]{Steid:96}
    Steidel C.C., Giavalisco M., Pettini M., Dickinson M., Adelberger K.L. 1996, \apjl, 462, L17
\bibitem[Tammann(1982)]{Tam}
    Tammann G. 1982, in: Supernova: A survey of current research,
    Eds. M. Ress and R. Stoneham, Dordrecht, Reidel, p. 371
\bibitem[Tscharnuter \& Winkler(1979)]{TW}
    Tscharnuter W.M., Winkler K.-H.A. 1979, 
    Comp.Phys.Comm. 19, 171
\bibitem[V\"olk(1987)]{Voe:87}
    V\"olk H.J. 1987,
    Proc.~20th Int.~Cosmic Ray Conf.~(Moscow) 7, 157
\bibitem[V\"olk et al.(1984)]{Voe:etal}
    V\"olk H.J., Drury L.O'C., McKenzie J.F. 1984, A\&A, 130, 19
\bibitem[V\"olk \& McKenzie(1981)]{Voe:mck}
    V\"olk H.J., McKenzie J.F. 1981,
    Proc.~17th Int.~Cosmic Ray Conf.~(Paris) 9, 246
\bibitem[V\"olk \& Zirakashvili(2004)]{VZ:04}
    V\"olk H.J, Zirakashvili, V.N. 2004, A\&A, 417, 807
%
%
\end{thebibliography}

\appendix

\section{Discretized equations}

The system of equations presented in Sect.~\ref{s.eqn} are rewritten to an
integral form, which is necessary for applying a finite volume discretization. 
Following this procedure ensures the numerical conservation of mass, momentum,
and energy. 

Introducing the temporal operator by $\delta x = x - x^{old}$ between the two time
levels $t= t^{old} + \delta t$, as well as the spatial difference $\Delta x_l = x_{l+1} - x_l$, 
we get a discrete version of the corresponding physical
equations (\ref{mass}) - (\ref{energy_w}). By defining the discrete volume
bet\-ween the distances $z_l$ and $z_{l+1}$ through
\begin{equation}  \label{e.dis_vol}
\Delta V_l = A_0 \left( \Delta z_l + \frac{\Delta z_l^3}{3 z_0^2}\right)
\end{equation}
we get, e.g.,
\begin{equation}  \label{e.dis_A}
 \pop{\Delta V_l}{z_{l+1}} = A_{l+1} = A_0 \left(1 + \frac{z_{l+1}^2}{z_0^2}\right)  \:.
\end{equation} 
Since the adaptive grid changes the locations of grid points, 
we have to compute the relative velocity 
\begin{equation}  \label{e.dis_urel}
   \urel_l = u_l - \frac{\delta z_l}{\delta t}
\end{equation}
with respect to the grid motion and $\urel$ appears in all advection terms.   

Written in a conservation form for a the flux-tube geometry, 
the discrete equation of continuity (\ref{mass}) reads as
\begin{equation}  \label{e.dis_cont}
\delta\left( \rho_l \DV_l \right) / \delta t
     + \Delta\left( A_l \urel_l \tilde \rho_l \right) = 0  \:,
\end{equation}
where $\tilde\rho_l$ denotes the density upstream to the relative velocity $\urel_l$.
Owing to the mathematical properties of hyperbolic conservation laws,
only those discretization schemes are stable that take 
the direction of the flow into account(see e.g.~\cite{LeV} for all mathematical details). 
The equation
of motion (\ref{momentum}) is given by 
\begin{eqnarray}  \label{e.dis_mot}
 \delta\left( \rho_l u_l \DV_l \right) / \delta t
      &+& \Delta\left( A_l \urel_l \tilde \rho_l \tilde u_l \right)
       + \left(P_{{\rm g},l}+P_{{\rm c},l} + P_{{\rm w},l}\right) \Delta A_l \nonumber \\ 
       &+& \rho_l \, \DV_l \nabla\Phi_{\rm gal} - u_{{\rm Q},l}\DV_l= 0  \:,
\end{eqnarray}
where $u_{{\rm Q},l}$ stands for the artificial pressure term added to broaden shock fronts
over a finite length scale, typically $\lenq = 10^{-3}z$. The
gas energy density (\ref{energy_g}) reads in the volume-integrated discrete form
\begin{eqnarray}  \label{e.dis_e_gas}
\delta\left( E_{{\rm g},l} \DV_l \right) / \delta t
     &+& \Delta\left( A_l \urel_l \tilde E_{{\rm g},l} \right) 
     + P_{{\rm g},l}\, \Delta (A_l u_l)  \nonumber \\
  &-& \varepsilon_{{\rm Q},l}\rho_l\DV_l = 0  \: ,
\end{eqnarray}
which also includes the viscous energy dissipation term $\varepsilon_{\rm Q}$. 
The cosmic ray energy density (\ref{energy_c}) is given through
\begin{eqnarray}  \label{e.dis_e_cr}
 \delta\left( E_{{\rm c},l}\Delta V_l \right) / \delta t
     &+& \Delta\left( A_l \urel_l \tilde E_{{\rm c},l} \right) \nonumber \\
     &+& (\gc-1)E_{{\rm c},l}\, \Delta (A_l u_l) = 0  \: .
\end{eqnarray}
Finally, we have also to solve the equation of wave energy density (\ref{energy_w})
\begin{eqnarray}  \label{e.dis_e_wav}
\delta\left( E_{{\rm w},l} \Delta V_l \right) / \delta t
     &+& \Delta\left( A_l \urel_l \tilde E_{{\rm w},l} \right) \nonumber \\
     &+& (\gw-1)E_{{\rm w},l}\, \Delta (A_l u_l) = 0  \: .
\end{eqnarray}

This set of physical equations is augmented by the grid equation, which redistributes the
grid points within the computational domain, and all details about this
numerical technique can be found in \citet{DD}. Therefore, in total we obtain $6N$
discrete nonlinear algebraic equations for the unknowns 
$\rho_l$, $u_l$, $E_{{\rm g},l}$, $E_{{\rm c},l}$, $E_{{\rm w},l}$, and $z_l$ at every 
index $l$, where $N=300$ is the number of spatial grid points. 
The resulting algebraic system is
solved at every new time step by adopting a 
Newton-Raphson iteration procedure where the corresponding Jacobian matrix has 
to be calculated from the discrete set of equations (\ref{e.dis_cont})-(\ref{e.dis_e_wav}).  

Due to the implicit nature of our numerical scheme we can compute time-dependent as
well as stationary solutions, which can directly be compared to the
stationary solutions obtained from solving the ODEs from the time-independent 
equations \citep{DB91}. 
Furthermore, in the steady state case integrals of 
the equations can be derived and used for a numerical check of the code. Both
methods show that the local, as well as global differences between the
solution of the ODE system and the numerical scheme presented above are less
than $10^{-5}$ everywhere. This small difference remains because of the staggered mesh
representation of the physical quantities and the relative accuracy of the
Newton-iteration of $10^{-6}$.

\section{Artificial viscosity}

Shock fronts are treated by applying an artificial viscosity in
tensor form based on the work of \citet{TW}. 
Since we are using an adaptive
grid \citep{DD} the length scale of the broadening $\lenq$ can be specified according to
our physical needs, i.e.~$\lenq \ll l_{\rm CR} = \kappa/\us$, which defines
the length scale due to the particle acceleration. This way we can ensure that
particles are accelerated by the 1.-order Fermi mechanism because the shock front 
is seen as a discontinuity by the energetic particles.

The tensor formulation of the artificial viscosity requires some calculations
for the adopted flux tube geometry since the pressure tensor $Q$, the
viscous momentum transfer $u_{\rm Q}$ and the viscous energy dissipation
$\varepsilon_{\rm Q}$ have to be calculated. Without more details
\citep[see][]{TW} we state these quantities
\begin{eqnarray}
   {\bf Q} &=& \mu_{\rm Q} \left[ \sgu 
             - \frac{1}{3} \diver{u}\,{\bf I}\right]           \label{e.Q}  \\
   u_{\rm Q}        &=& -\diver{{\bf Q}}                       \label{e.uQ} \\
   \varepsilon_{\rm Q} &=& -\frac{1}{\rho}\,{\bf Q}\cdot\!\sgu, \label{e.eQ} 
   \end{eqnarray}
where $\sgu$ is the symmetrized velocity gradient and 
${\bf I}$~the unity tensor.

Artificial viscosity coefficient $\mu_{\rm Q}$ consists of
a linear and a quadratic term weighted by $q_1$ and $q_2$, i.e.,
\begin{equation}\label{e.muQ}
   \mu_{\rm Q} = - q_1 \lenq c_{\rm s} + q_2^2 \lenq^2 \min ( \diver{{\bf u}} , 0 ) .
\end{equation}
The definition of $\mu_{\rm Q}$ include a typical viscous length
scale $\lenq$. When inspecting the last expression one sees that the linear term in $\lenq$ 
also scales with the sound velocity $c_{\rm s}$ and is always present 
when $q_1 \ne 0$. The second term is quadratic in $\lenq$, 
and it is evident that {\em compressive and nonhomologous} motions; 
i.e., $\diver{\bf u}<0$ and $tr({\bf Q}) \ne 0$ produce a viscous pressure.
The amount of artificial viscosity is then determined by these length scales
$q_1 \lenq$ and $q_2 \lenq$. 

To obtain the corresponding expressions (\ref{e.uQ}) and (\ref{e.eQ}) for 
our flux tube geometry, we recall that the distance
from the galactic plane is denoted by $z$ and the area of the
flux tube is defined by $A(z)$. Neglecting the curvature within the 
flux tube area, we get the simple orthogonal metric tensor by
\begin{equation} \label{e.gik}
  g_{ik} =\left(\begin{array}{ccc}
               1 &   0 & 0   \\
               0 & z^2 & 0   \\
               0 &   0 & A^2 \\ 
              \end{array}\right)  \: .
\end{equation}
Following the rules derived by \citet{TW},
we can calculate the divergence of the velocity from the last expression
\begin{equation}  \label{e.divu}
  \diver{{\bf u}} = \frac{1}{A}\pop{Au}{z} = \pop{Au}{V} \: ,
\end{equation}
as well as the symmetrized velocity gradient
\begin{equation}  \label{e.eps_lm}
   \epsilon_l^m =\left(\begin{array}{ccc}
               \displaystyle{\pop{u}{z}} & 0                   & 0 \\
               0       & \displaystyle{\frac{u}{2A}\pop{A}{z}} & 0 \\
               0       & 0   & \displaystyle{\frac{u}{2A}\pop{A}{z}} \\ 
              \end{array}\right)  \: .
\end{equation}
The viscous pressure tensor is given through
\begin{eqnarray} \label{e.p_visc}
   {\bf Q} &=& \frac{1}{A}\pop{Au}{z} \ \times \\ 
    &\times& \left( \begin{array}{ccc}
            \displaystyle{\pop{u}{z} - \frac{1}{3A}\pop{Au}{z} } & 0 & 0 \\
            0 & \displaystyle{\frac{u}{2A}\pop{A}{z}- \frac{1}{3A}\pop{Au}{z} } & 0 \\
            0 & 0   & \displaystyle{\frac{u}{2A}\pop{A}{z}-\frac{1}{3A}\pop{Au}{z} } \\ 
                   \end{array} \right) \nonumber  
\end{eqnarray}
and has the desired property of being traceless; i.e., there is no artificial
viscosity in the case of homologous motions. 
According to the previous terms, the viscous pressure is given in the flux tube
geometry through
\begin{equation}
  u_{\rm Q} = \frac{1}{A^{3/2}} \pop{}{z} \left[ A^{1/2} \mu_{\rm Q}\,\rho
     \pop{Au}{z}\left(\pop{u}{z}- \frac{1}{3}\pop{Au}{z}\right) \right]\:.
\end{equation}
The energy dissipation by the viscosity is determined from the contraction of the
pressure tensor with the symmetrized velocity gradient (see Eq.~\ref{e.eQ}).
To ensure physical meaningful results it is necessary to find a discrete version,  
which always guarantees a positive viscous energy 
generation. This term has been implemented in our code through
\begin{equation}
   \varepsilon_{\rm Q} = - \frac{3}{2}\frac{\mu_{\rm Q}}{A} \pop{Au}{z} 
                           \left[\pop{u}{z} - \frac{1}{3A}\pop{Au}{z}\right]^2  \: , 
\end{equation}
guaranteeing that the viscous energy generation  $\varepsilon_{\rm Q}$
is also numerically positive.

According to the discretization scheme, the artificial viscosity
terms are written as
\begin{equation}
   \mu_{{\rm Q},l} = q_1 \lenq\, c_{{\rm s},l} 
        - q_2 \lenq^2 
          \min\left(\frac{\Delta(z_l^2 u_l)}{\DV_l},\, 0 \right) , 
\end{equation}
including the linear part with $c_{{\rm s},l}$, the 
adiabatic sound velocity at a grid point $z_l$. The corresponding discrete
and volume-integrated versions of $u_{\rm Q}$ and $\varepsilon_{\rm Q}$ 
are then given as follows.   
The viscous pressure term in the discrete equation of motion 
(\ref{e.dis_mot}) reads as
\begin{eqnarray}
   u_{{\rm Q},l}\DV_l &=&    \\
   &-& \frac{1}{\sqrt{A_l}}\Delta
       \left( \bar A^{\,3/2}_l \mu_{{\rm Q},l}\,\rho_l \frac{\Delta(A_lu_l)}{\DV_l}
          \left[ \frac{\Delta u_l}{\Delta z_l} 
                 - \Zdri\frac{\Delta (A_lu_l)}{\DV_l} \right] \right) \nonumber \: ,
\end{eqnarray}
and the corresponding viscous energy dissipation (\ref{e.dis_e_gas}) 
is discretized at a grid point $z_l$ according to 
\begin{equation}
   \varepsilon_{{\rm Q},l}\rho_l\DV_l =  \frac{3}{2} \mu_{{\rm Q},l}\, \rho_l \Delta ((A_lu_l))
    \left[\frac{\Delta u_l}{\Delta z_l}-\Zdri\frac{\Delta A_lu_l}{\DV_l}\right]^2\DV_l \:.
\end{equation}

\end{document}